# 3D and 4D printing in dentistry and maxillofacial surgery: Recent advances and future perspectives

Danial Khorsandi[1,2,3], Amir Fahimipour[4], Payam Abasian[5], Sepehr Sadeghpour Saber[6], Mahla Seyedi[7], Sonia Ghanavati[8], Amir Ahmad[6], Andrea Amoretti De Stephanis[6], Fatemeh Taghavinezhad[6], Anna Leonova[9], Reza Mohammadinejad[10], Majid Shabani[11,12], Barbara Mazzolai[11], Virgilio Mattoli[11], Franklin R. Tay[13*], Pooyan Makvandi[11*]

[1]Department of Biotechnology-Biomedicine, University of Barcelona, Barcelona 08028, Spain
[2]Department of Mechanical engineering, McMaster University, Hamilton, ON L8S 4L8, Canada
[3]Department of Medical and Scientific Affair, Procare Health Iberia, Barcelona 08860, Spain

[4]Discipline of Oral Surgery, Medicine and Diagnostics, School of Dentistry, Faculty of Medicine and Health, Westmead Centre for Oral Health, The University of Sydney, NSW 2145, Australia

[5]Department of Chemical Engineering, Isfahan University of Technology, Isfahan, Iran

[6]International University of Catalunya, Faculty of Dentistry, Sant Cugat, Barcelona 08195, Spain

[7]Department of Engineering, Corrosion and Metallurgy Study Center "A. Dacco", University of Ferrara, Ferrara 44121, Italy

[8]Department of Medicine and Health Technology, Tampere University of technology, Tampere 33720, Finland

[9]Department of Medical Sciences, McMaster University, Hamilton L8S 4L8, ON, Canada

[10]Research Center for Tropical and Infectious Diseases, Kerman University of Medical Sciences, Kerman 76169-13555, Iran

[11]Istituto Italiano di Tecnologia, Centre for Micro-BioRobotics, viale Rinaldo Piaggio 34, 56025 Pontedera, Pisa, Italy

[12] The BioRobotics Institute, Scuola Superiore Sant'Anna, viale Rinaldo Piaggio 34, 56025 Pontedera, Pisa, Italy

[13]The Graduate School, Augusta University, Augusta, Georgia 30912, United States

Corresponding authors:
Email: ftay@augusta.edu (F. Tay); Pooyan.makvandi@iit.it (P. Makvandi)

**Abstract**

3D and 4D printing are cutting-edge technologies for precise and expedited manufacturing of objects ranging from plastic to metal. Recent advances in 3D and 4D printing technologies in dentistry and maxillofacial surgery enable dentists to custom design and print surgical drill guides, temporary and permanent crowns and bridges, orthodontic appliances and orthotics, implants, mouthguards for drug delivery. In the present review, different 3D printing technologies available for use in dentistry are highlighted together with a critique on the materials available for printing. Recent reports of the application of these printed platformed are highlighted to enable readers appreciate the progress in 3D/4D printing in dentistry.

**Table of Contents**

**Abbreviations**

2,2´-azo-bis-isobutyrylnitrile (AIBN); Acrylonitrile butyrostyrene (ABS); Acrylonitrile butyrostyrene (ABS); Additive manufacturing (AM); Alumina toughened zirconia (ATZ); Alumina trihydrate (ATH); Aluminium ($Al_2O_3$); Ceria-stabilized zirconia (12Ce-TZP); Chromium oxide ($Cr_2O_3$); Cobalt chromium molybdenum (CoCrMo); Commercially pure titanium (CP-Ti); Cone beam computed tomography (CBCT); Controlled radical polymerization (CRP); Digital light projection (DLP); Digital micromirror device (MDM); Energy dispersive X-Ray (EDX); Fused deposition modeling (FDM); Fused filament fabrication (FFF); Free radical polymerization (FRP); Hexagonal close-packed (HCP); Magnesium oxide (MgO); n-butyllithium (n-BuLi); Polylactic acid (PLA); Poly(methyl methacrylate) (PMMA); Polycaprolactone (PCL); Polycarbonate (PC); Polyetheretherketone (PEEK); Polystyrene (PS); Polyvinylsiloxane (PVS); Rapid prototyping (RP); Scanning electron microscopy (SEM); Selective laser sintering (SLS); Stereolithography (SL); Stereolithography apparatus (SLA); Temporomandibular joint (TMJ); Titanium dioxide ($TiO_2$); Ultraviolet (UV); Zirconia-toughened alumina (ZTA); Zirconium oxide ($ZrO_2$)

**1. Introduction**

Three dimensional (3D) printing is an industrial technology that has rapidly evolved over its forty-year history [1]. This additive manufacturing (AM) approach differs from classical subtractive manufacturing principles and is currently utilized in a plethora of disciplines ranging from aerospace industries to personalized medicine and dentistry. This manufacturing scheme enables rapid creation of custom-based complex parts, made it an applicable solution in developing self-growing robots [2]. Additive manufacturing is represented by technologies such as stereolithography, fused deposition modeling, selective laser sintering, inkjet printing, photopolymer jetting and powder binder printing [3–8]. Different materials, e.g., polymers, composites, ceramics, and metal alloys, are employed for additive manufacturing [9,10]. Integration of 3D printing into different facets of contemporary dentistry has enabled the production of complex prosthodontic, orthodontic and surgical devices that demand flexibility and abrasion resistance from the molding materials.

Recent progress in 3D-printable smart materials has brought into fruition a newer generation of "dimensional printing" that is coined 4D printing. Four-dimensional printing is the combination of 3D printing with time as the $4^{th}$ dimension [11]. Such a printing platform produces pre-programable bio-objects that change their shape in response to the surrounding media [12].

Companies are taking advantage of recent expiration of the initial patents to introduce and expand their technologies (**Figure 1**). In light of 3D and 4D printing rapidly becoming a manufacturing and prototyping alternative in dentistry, the objective of the present review is to explore, explain and compare the different 3D printing technologies that have been utilized for the fabrication of dental biomaterials. Applications of these 3D printed platforms will also be described to enable readers appreciate the progress achieved in this rapidly developing field.

**Table 1.** Advantages and disadvantages of 3D/4D printing methods in dentistry.

| 3D/ 4D printing system | Material characteristics | Materials | Advantages | Disadvantages | Ref. |
|---|---|---|---|---|---|
| Stereolithography (SLA) | Light curable resin | Epoxy and methacrylate monomers | - Product resolution<br>- Efficiency<br>- Short working time | - Over-curing<br>- Lack of surface smoothness<br>- Limited mechanical strength<br>- Irritant | [13–16] |
| Selective laser sintering (SLS) | Powder | - Polymers<br>- Ceramics<br>- Metals | - Structures are fully self-supporting<br>- Protective gas in not needed<br>- Vast variety of materials can be selected<br>- Little to no thermal stresses are accumulated on the component<br>- Components exhibit excellent mechanical properties<br>- Relatively fast method | - Sample surfaces appear porous and rough<br>- Harmful gases release during fabrication<br>- Materials waste is relatively high<br>- Raw powders are expensive to an extent<br>- Post-processing is often expensive and tedious | [17–20] |
| Fused deposition modeling (FDM) | Thermoplastic polymer and composites, low melting temperature metal alloys | - Paste<br>- Wire | - Filaments are cheap and arrive in various colors<br>- Easy to change materials<br>- Cost-effective maintenance<br>- Capable of fast production of shelled structures<br>- Fundamental for thinner layers up to 0.1 mm thick<br>- Released fumes are not toxic | - The seam between layers is visible<br>- Discontinuous extrusion results in formation of defects<br>- Support structure is required in some cases<br>- Delamination between layers may occur due to low extrusion temperature<br>- Printed component may curl off the build platform because of induced thermal stresses | [9,21–25] |
| Photopolymer jetting | Light curable resin | - Biocompatible (MED610)<br>- VeroDentPlus (MED690) and VeroDent (MED670)( all are natural looking medically - approved photopolymers) | - High resolution due to thin layer printing (~16 microns per layer)<br>- Short working time<br>- Excellent surface features<br>- No need for post-modification<br>- Supporting wide range of materials | - Irritant<br>- High cost | [26–32] |

| Powder binders | Materials which are available in powder | - Metal<br>- Ceramic<br>- Plastics | - Safe material<br>- Short working time<br>- Suitable mechanical performance<br>- Low cost | - Low resolution<br>- Low strength<br>- Cannot be soaked/heat sterilized | [33,34] |
|---|---|---|---|---|---|
| Digital light projection | Light curable resin | - Resin | - High complexity and excellent surface finish<br>- Short timeframe<br>- Good accuracy<br>- Smooth surfaces | - Limited material selection<br>- Photocurable resin can cause skin sensitization, and maybe irritant by contact | [35] |
| Computed axial lithography | Light curable resin | - Resin containing dissolved oxygen | - It can be used in specific conditions where existing methods fall short, such as:<br>. Printing soft materials that cannot maintain the forces applied during layerwise printing,<br>. Creating lenses with smooth curved surfaces,<br>. Encapsulating other objects in three dimensions | - Limited material selection | [36,37] |

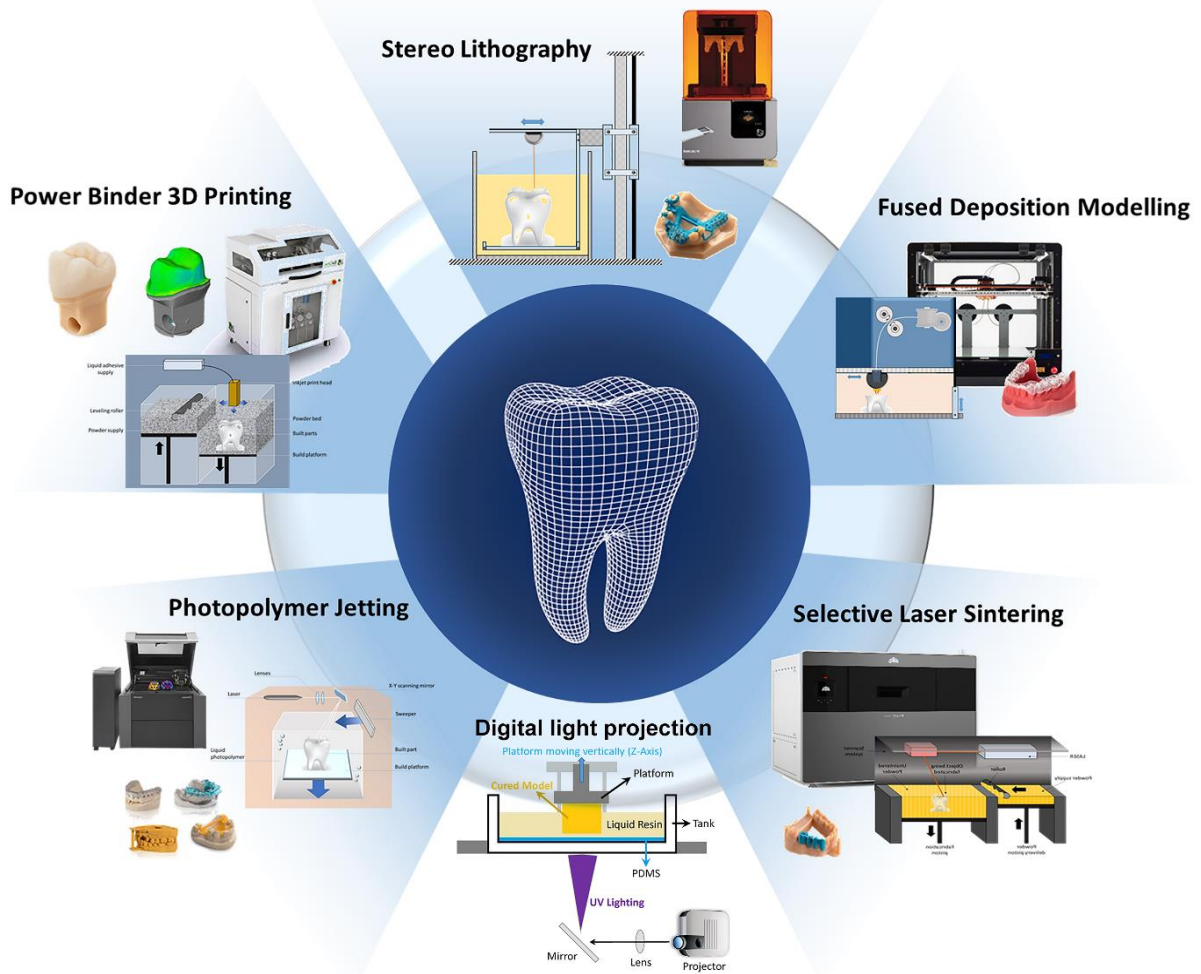

**Figure 1.** Current 3D and/or 4D printing techniques used in dentistry.

## 2. 3D printing techniques

### 2.1 Stereolithography

Stereolithography (SLA) is the first commercially available 3D printing technology. This rapid manufacturing method involves photoinduced polymerization to create layered structures using highly cross-linked polymers[38]. This technology may be subdivided into different categories based on the type of platform motion and laser movement. Irrespective of these categories, printing proceeds through three major steps: light/laser exposure, platform movement, and resin refilling. Stereolithography is the typical representative of AM, which utilizes layer-by-layer modeling. A 3D digitized model, which is used as the template for the fabrication process, guides the SLA machinery to complete the printed object. The layers are bound together bottom-up upon exposure of the resin to ultraviolet light, which induces free radical polymerization (FRP) of the resin monomers. As one layer is polymerized, the resin platform lowers by a distance equal to the thickness of one layer and builds the next layer until the printing of the digitized 3D object is completed (**Figure**

**2A** and **B**) [39]. The discontinued manner of processing can be overcome by combining SLA with continuous liquid interface production. The latter a proprietary method of 3D printing that uses photopolymerization to create smooth-sided solid objects [25,40] Variables such as light source intensity, scanning speed as well as the amount of resin monomers and photoinitiators may be controlled to achieve the required modeling kinetics and properties of the final product [31]. Currently, SLA is applied in the manufacturing of temporary and permanent crowns and bridges, temporary restorations, surgical guides, templates, and dental model replicas (**Figure 2B**) [26]. **Table 1** represents the pros and cons of 3D/4D printing methods in dentistry and maxillofacial surgery.

Stereolithography accommodates flexibility in design, geometric shape and scaling, resulting in highly accurate personalized devices. High precision measurements retrieved from the patient's scanning data enable production of reliable appliances for long-term use [41]. However, cytotoxicity of the printed appliances may be caused by the leaching of residual unreacted resin monomers from the printed appliances. The may affect the longevity of the appliances.

Stereolithography demonstrated greater clinical accuracy than other digital/analog methods in the production of dental stone casts; the 3D printing technique represents an acceptable alternative for diagnosis, treatment, and production of prosthetic devices [42]. Nevertheless, SLA-printed dental devices suffer from poor mechanical properties caused by the limited choice of resins that can be photopolymerized [39]. Nanoparticle incorporation into the polymeric matrix improves mechanical properties [43]. Other coupling agents such as ceramic fillers protect the printed structure from fracture by improving stress distribution. Antimicrobial agents have also been incorporated into the resins to address the issue of microbial colonization of oral devices [14,44]. Overall, SLA is a rapid, convenient and multifunctional technique in 3D dental printing (**Figure 2**).

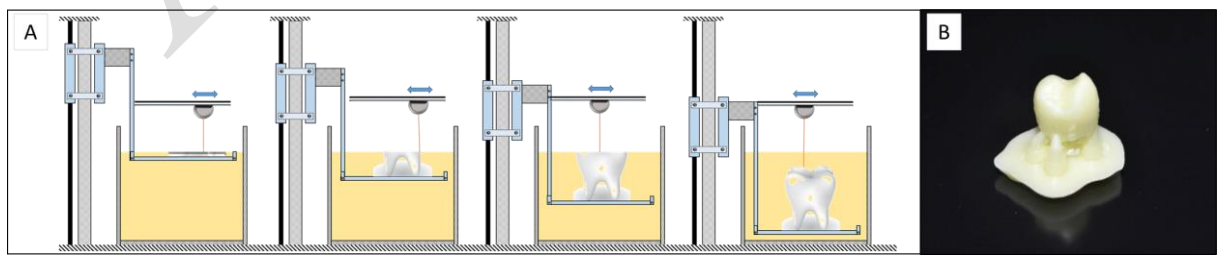

**Figure 2.** Stereolithography printing. (A) Layer by layer printing process. (B) An example of a crown printed with stereolithography prior to removal of supports and polishing.

## 2.2 Digital light projection

DLP is a photocuring technology which is similar to SLA process. The materials are liquid photosensitive resins which undergo photocuring and subsequently form the 3D printed part layer by layer. The first layer is formed on the build platform. Based on the position of the UV source, the build platform may be ascending or descending as shown in **Figure 3B** [45]. Next layers will be formed on their previous layers. DLP 3D printer utilizes a digital projector screen to flash the current layer's image, through a transparent bottom/top of the resin tank, across the build platform or previous layer. After curing each layer, the build platform goes up/down as the thickness of a layer until completing the entire part [46,47]. A digital micromirror device (MDM) is used to reflect the light. DMD consists of a matrix of microscopic-size mirrors. These mirrors conduct the light from the laser projector to the projection lens. They make different configurations, adjustable for each layer, such that they create the 2D sketch of the layer by light on the curing surface [48].

Although SLA and DPL are very similar, they also have some differences. The main difference is the light source. SLA benefits from UV laser beam while DLP uses UV light from the projection source. Consequently, in SLA laser beam moves from point to point and cures the resin from point to point while in DLP the light source is stationary and cures each layer of the resin at a time. These different curing processes result in more accurate and better quality in SLA compared to DLP, while on the other hand improves the printing speed in DLP method. The intensity of the light source in the DLP 3D printer is adjustable while it is not adjustable in the SLA printer. This means that the operator can control the effect of light on the resin. In summary, DLP is advantageous in the fast printing of bigger parts with fewer details while SLA is advantageous in printing accurate parts with intricate details [49–53].

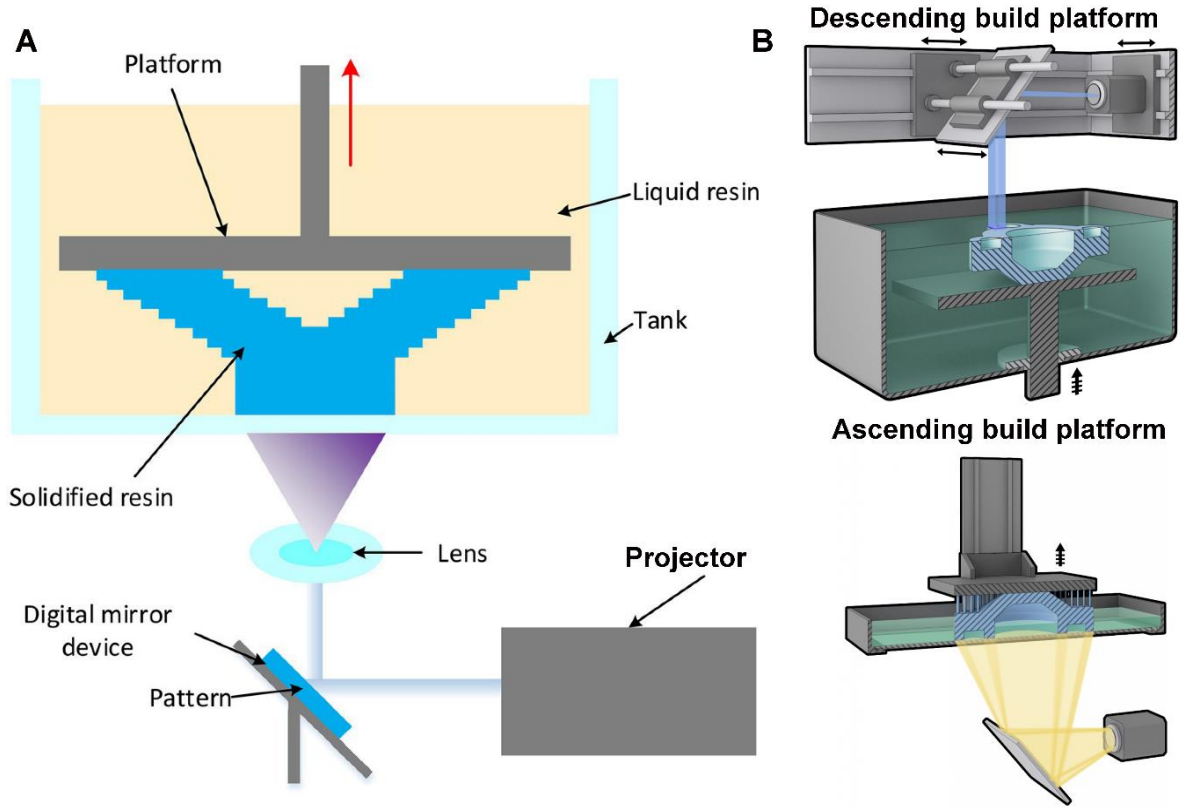

**Figure 3.** Digital light projection (DLP). (A) Schematic of the printer components and printing procedure. Reprinnted with permission from [49]. (B) Demonstration of two kinds of platform movements in DLP method [54].

## 2.3 Fused deposition modeling

Fused deposition modeling (FDM), also known as fused filament fabrication (FFF), is the trade name for a polymer, composites, or metal alloys softening process that was invented over 20 years ago. It is the second most commonly used 3D printing technique, after SLA [55]. This method is remarkably cheaper than the other AM techniques [56]. The basis of this technique is the principle of strand extrusion: the desired type of thermoplastic materials, shaped as strands, is delivered to the extruder. Upon softening, the heated viscous plastic is deposited by an extrusion head that results in layer-by-layer fabrication of the digitized model (**Figure 4A**) [57,58]. Unlike SLA, individual layers within the object have reduced bonding and thus, the final product has greater anisotropy [25]. Thermoplastic polymers and their composites (e.g., acrylonitrile-butadiene-styrene (ABS), polycarbonates and polysulfones) along with low melting temperature metal alloys (e.g., bronze metal filament) are the most common employed FDM filaments [9,56]. Polymers may be filled with metal (nano)particle reinforcement to prepare (nano)composite filament to improve different features, e.g., thermal resistance and mechanical properties. Mechanical properties in the FDM method can be affected by three main groups of parameters: print material, structural parameters (i.e. rasters angle, infill density, print orientation, and stacking sequence), and manufacturing parameters (i.e. extrusion temperature and rate, layer time, nozzle transverse speed, and bed temperature) [56]. Because of the weak mechanical properties of these unfilled thermoplastics, FDM is used only for the printing of temporary crowns and bridges in dentistry. Similarly to SLA, processing commences with the acquisition of computer-aided design (CAD) images, as depicted in **Figure 4B**. Based on these digitized images, FDM is capable of rapid printing of the destined product (**Figure 4C**).

Fused deposition modeling is the technology utilized in most low-cost 'home' 3D printers. It enables the printing of crude anatomical models without too much complexity. Although the final product is brittle and contains rough surfaces, FDM is cheaper than other 3D printing techniques. This technique is used in machines with low maintenance and in scientific research [55].

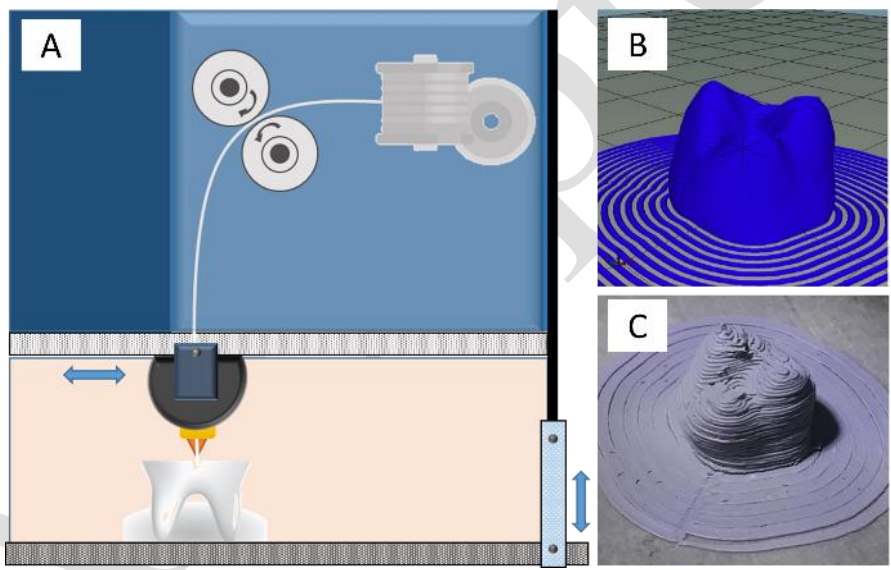

**Figure 4.** Fused deposition modeling (FDM). (A) FDM printer. (B) dental crown CAD file. (C) dental crown FDM. Reprinted with permission from [59].

## 2.4 Selective laser sintering

Selective laser sintering (SLS) is a system wherein a high energy beam laser is used to induce fusion of the powdered raw material. The laser creates a solid layer out of the powder, The platform will be lowered to make space for the laser to sinter the next layer of powder. This method of structure creation does not require additional material support during printing because support is provided by the powder surrounding it [60].

Multi-purpose study models, guides for drilling and cutting as well as metal frameworks can be created using SLS (**Figure 5**). The advantages of SLS include the use of autoclavable materials, mechanical functionality of the printed object, reduction in production cost with increase in production volume. The disadvantages of this printing technique are the health risk associated with inhalation of the powdered raw material, the initial high cost in setting up and the need for supplementary supplies such as compressed air for proper functioning of SLS [17,19,20]. The function of the laser is to increase the temperature of the powder to close to, but not its sintering point. This sintering process converts the solid power into a semi-liquid state. The platform on which the first layer rests lowers by ~0.1 mm, giving the laser space for a new layer of powder to be sintered. This stepwise sintering and fusion continue until the object is fully printed. The object is left to cool down after printing is complete [19].

Another benefit of SLS is the almost utilization readiness of the printed object. Whereas other printer methods necessitate an extra sanding step or other forms of finishing before the printed products are usable, this step is usually not mandatory for the SLS technique. Because SLS printers do not require structural support during printing, the processing time is considerably faster compared with SLA and FDM [61–63]. Three types of SLS printing are available: metal-based, ceramic-based, and polymer-based SLS. The metal-based SLS technique utilizes fine metal powders, whereas the polymer-based SLS technique utilizes fine-grained thermoplastic polymers as raw materials. For metal-based SLS, different metal powders may be employed, such as stainless steel alloys (316, 304L, 309, 174), and titanium alloy (6AI-4V). Unlike polymer-based SLS which does not require gas injection, metal-based SLS requires protective gas to be flooded into the printing chamber to avoid oxidation of the metal powder when the latter is heated to a high sintering temperature.

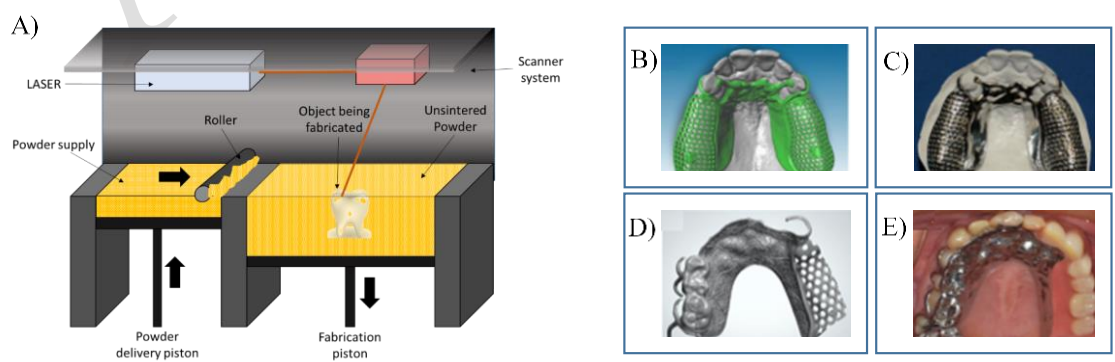

**Figure 5.** Selective laser sintering (SLS). (A) illustrations of the design and fabrication of a removable partial denture (RPD) framework by SLS. (B) CAD design of RPD framework, (C) the final RPD framework, (D)

laser-sintered metal frame, (E) partial denture fabricated using the sintered framework. Reprinted with permission from [64].

In addition, SLS technique has been employed for fabricating ceramic parts [65]. Since ceramic powders need appropriate exposure time to result in the desired density, the target temperature should be lowered to facilitate the densification. One solution is mixing other materials such as organic (e.g., polymers) or inorganic (e.g., metal-based low-melting materials and glass) compounds, as binders in the ceramic powder. Therefore, binders melt in the heated powder bed surface and creat a bonding glassy phase around the ceramic particles. Besides, these binders are more sturdy to temperature variations. To prevent the oxidation of the binder contents, insert atmospheres is mandatory. Organic contents can be removed in the high-temperature firing in furnace, while the inorganic parts remain [66]. Infiltration/isostatic pressing along with SLS enhances the mechanical performance of the ceramic parts by maximizing the final density [67]. However, there are several challenges in the fabrication of ceramic parts with optimal mechanical performances. Powder deposition, the interaction between laser and powder, mechanism of melting and consolidation, as well as thermal and residual stress analysis are some of these challenges [68].

## 2.5 Photopolymer Jetting

This technology involves the combination of two techniques, using a dynamic printing head and photopolymerizable polymer. The light-sensitive polymer is jetted to a building platform from an inkjet-type printing head and cured layer by layer on the descending platform (**Figure 6**) [69]. A supporting structure is also printed using fragile support material for easy removal. This technology enables printing of an extensive range of resins and waxes for casting. Silicone-like rubber materials may also be used for printing complex and highly-detailed products with a resolution of ~16 microns. The printed products may be used as crowns or anatomical study models. Another advantage of the printing technique is that implant drill guides may be printed rapidly and economically with better quality. 3D jet printers utilize multiple printing heads to cover the working platform width [27,30].

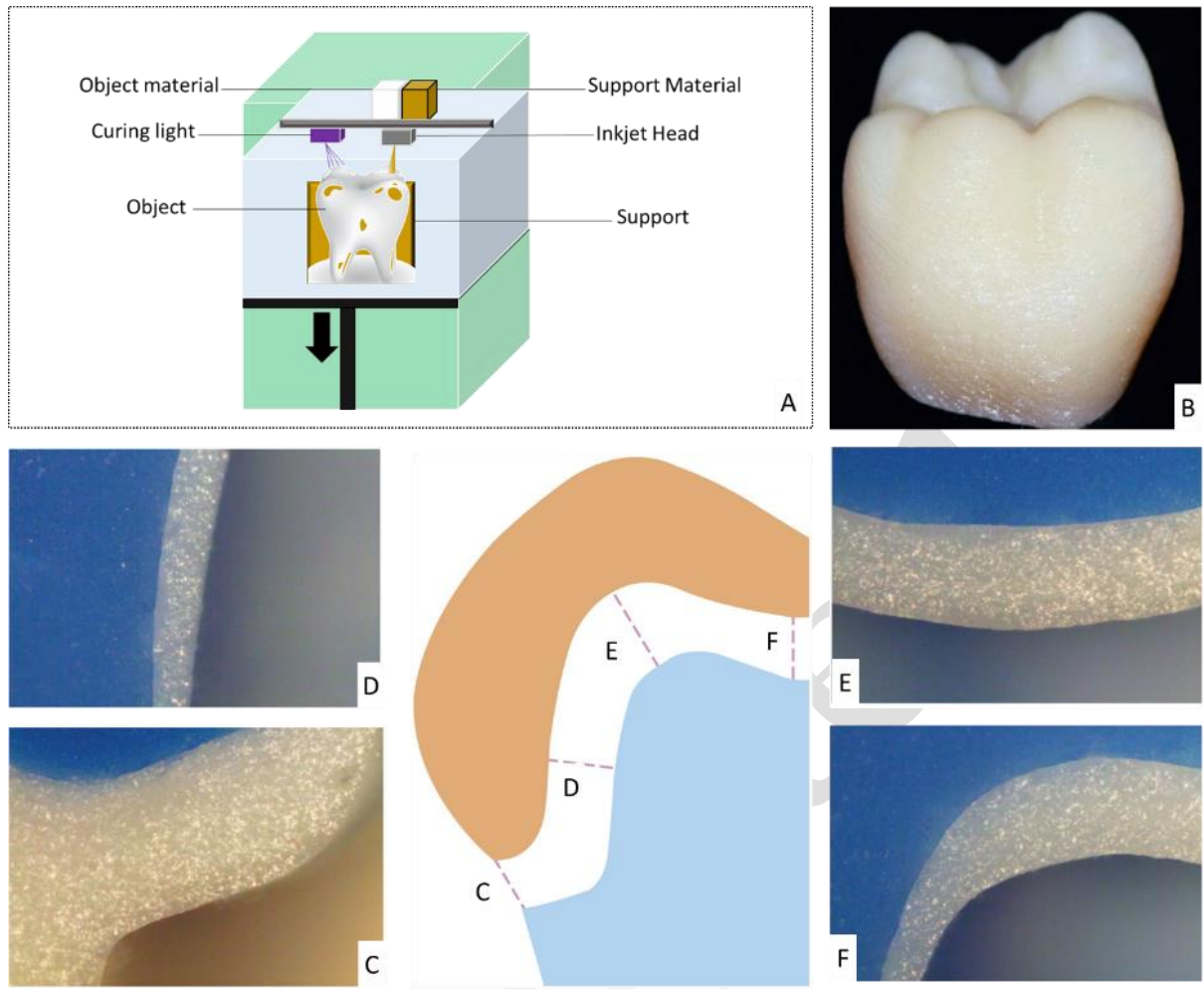

**Figure 6.** (A) Photopolymer jetting. (B) Temporary crowns fabricated with the photopolymer jetting method. Micrographs of the cross-sections of the temporary crown (C) Margin. (D) Axial. (C) Cusp. (D) Fossa. (Magnification ×50). Reprinted with permission from [27].

The printing head and working platform move bidirectionally but independently. An ultraviolet light source is used to harden each layer of resin or wax that has been jetted. The advantages of this technology are rapid manufacturing, surface smoothness and cost efficacy. The drawbacks include difficulty in removing the material completely due to rigid support, skin irritation, inability to be heat-sterilized and the high cost of the material [25,26].

## 2.6 Powder Binder Printer

This AM process uses liquid adhesive droplets within a modified inkjet head. The inkjet head releases these droplets for infiltrating a layer of powder located underneath. This step is repeated by replenishing a new layer of powder until the final product is manufactured. The main application of powder binder printing in dentistry is in the printing of study casts or prototypes. However, the printed objects are fragile and lacked accuracy. Because this is low-cost technology, the technique finds use in applications that do not require sterilization, such as printing of study casts. However, the manufacturing process is messy because of the use of powder (**Figure 6**) [70].

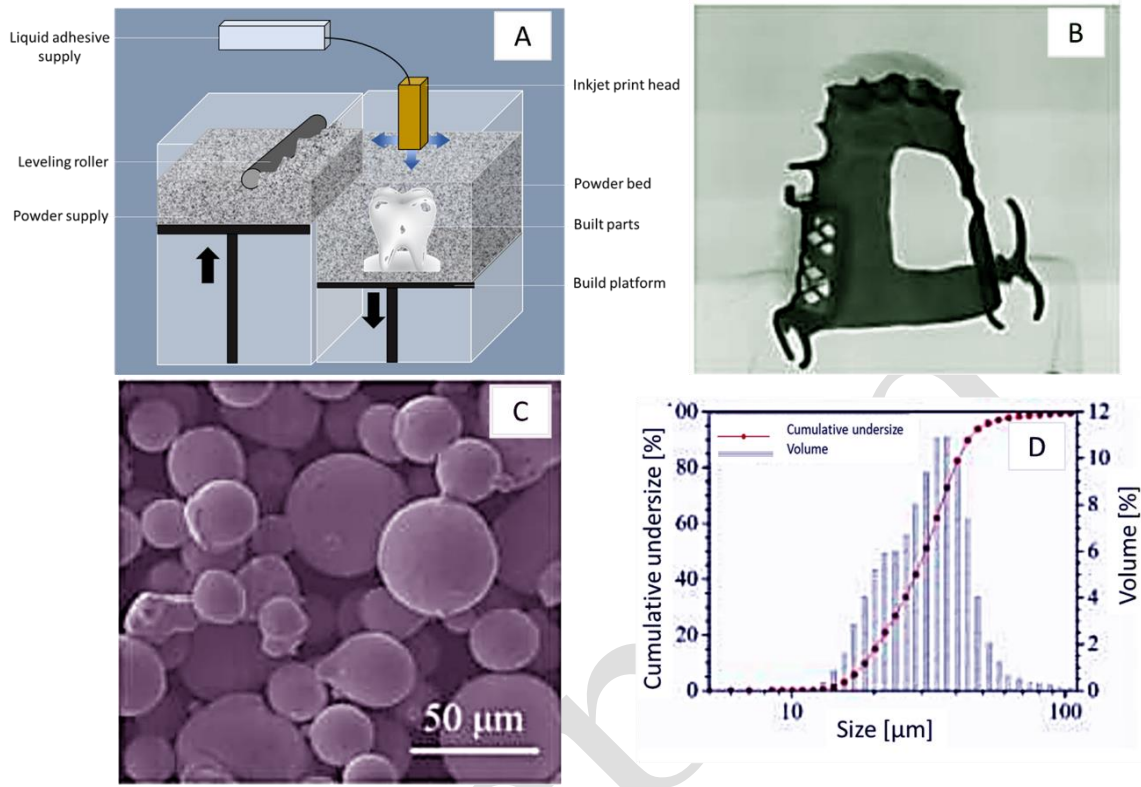

**Figure 7**. (A) Powder binder printing. (B) A 3D printed denture model. (C) Scanning electron micrograph of the powder used for printing. (D) Distribution in particle size of the powder employed for printing**.** Reprinted with permission from [71].

## 2.7 Computed axial lithography

There is intense interest in the adaptation of novel technologies for 3D/4D printing in dentistry. Volumetric AM is an example of these newly developed systems. Computed axial lithography is a form of volumetric AM printing technique. It has some similarity to DLP method as both systesm employ projector as light source to photopolymerize the resin. The difference between computed axial lithography and other conventional printing technqiues is that light polymerization is applied at several angles to the material, where as other AM printing methods utilize layer-by-layer polymerization. In other words, computed axial lithography can produce the entire object at once (not layer by layer). The idea of computed axial lithography is inspired by computed tomography scanners. In computed tomography, X-ray scanning is performed from many different angles. This concept was utilized in the development of computed axial lithography [35]. The object is fabricated by projecting a light (with specific wavelength) containing many 2D pictures (from different angles of the object) to a rotating container of photocurable resin (**Figure 8**). Photopolymerization of the resin material permits manufacturing of products with higher complexity and better surface finish in shorter timeframe, compared with other AM methods. Centimeter-scaled products may be printed within a minute with the use of computed axial lithography. The products printed with this technique can be easily designed with CAD and saved in stereolithography file format. To date, high viscosity resins or solid materials are used predominantly for this technique. The application of materials with low viscosity may permit prining of tissues in the future. However, more *in vivo* studies are required to investigate the mechanical and chemical properties, tolerability and biocompatibility of the photopolymerized resin products prior to their recommendation for clinical applications [35,72].

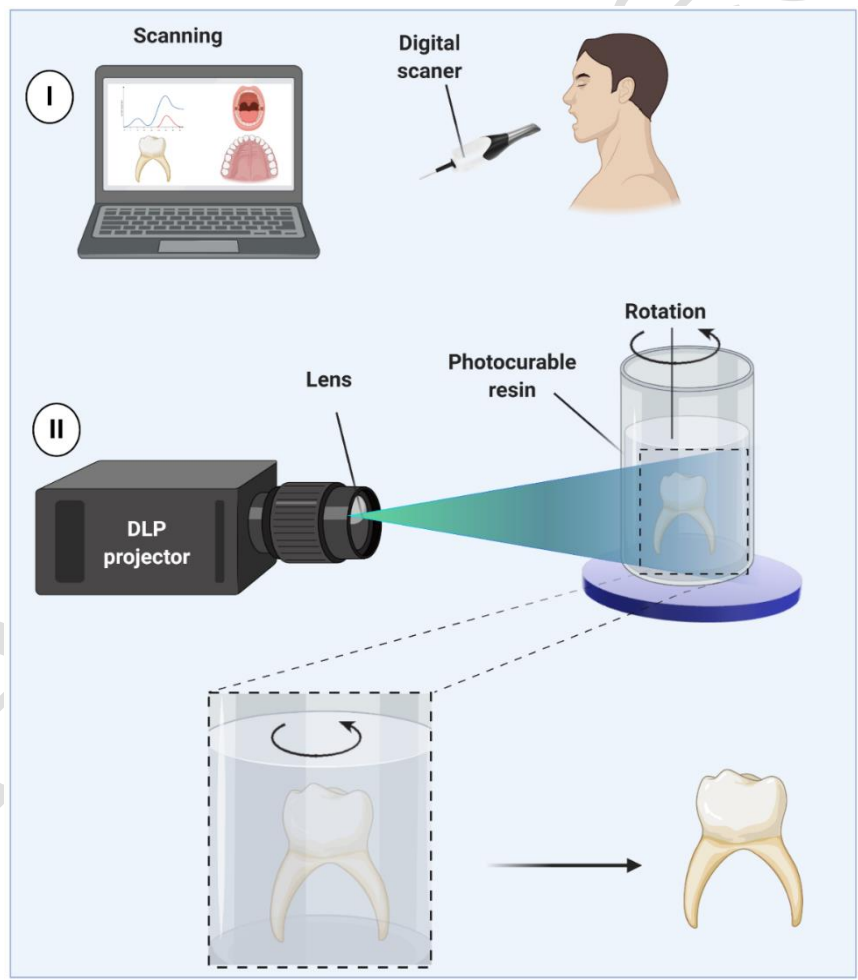

**Figure 8.** Schematic of scanning (I) and then printing (II) using computed axial lithography as volumetric additive manufacturing. A 3D model is initially prepared using scanners (I). The 3D model is converted into many 2D images. Each image is projected in one angle to the rotating container. DLP: digital light processing.

## 3. Materials employed for 3D/4D printing in dentistry

### 3.1 Polymers

*3.1.1 Vinyl polymers*

Vinyl polymers are the most commonly used polymers in dentistry because of their tunable properties. They are produced from vinyl monomers using FRP (**Figure 9A**). Although vinyl polymers are biocompatible, the majority of them are not biodegradable making them unfavoured material for many medical applications, but not dentistry, since degradation is not desirable for a long-term use, such as dental implants. For this reason, new synthesis methods such as controlled radical polymerization (CRP) have been implemented to tune material properties and control the molecular weight of the resulting polymer, as well as their end and chain functional groups. Vinyl polymers are extensively used for dental 3D printing that involves sintering (e.g., SLS) or photopolymerization (e.g., SLA) **(Figure 9B)** [70,73–76].

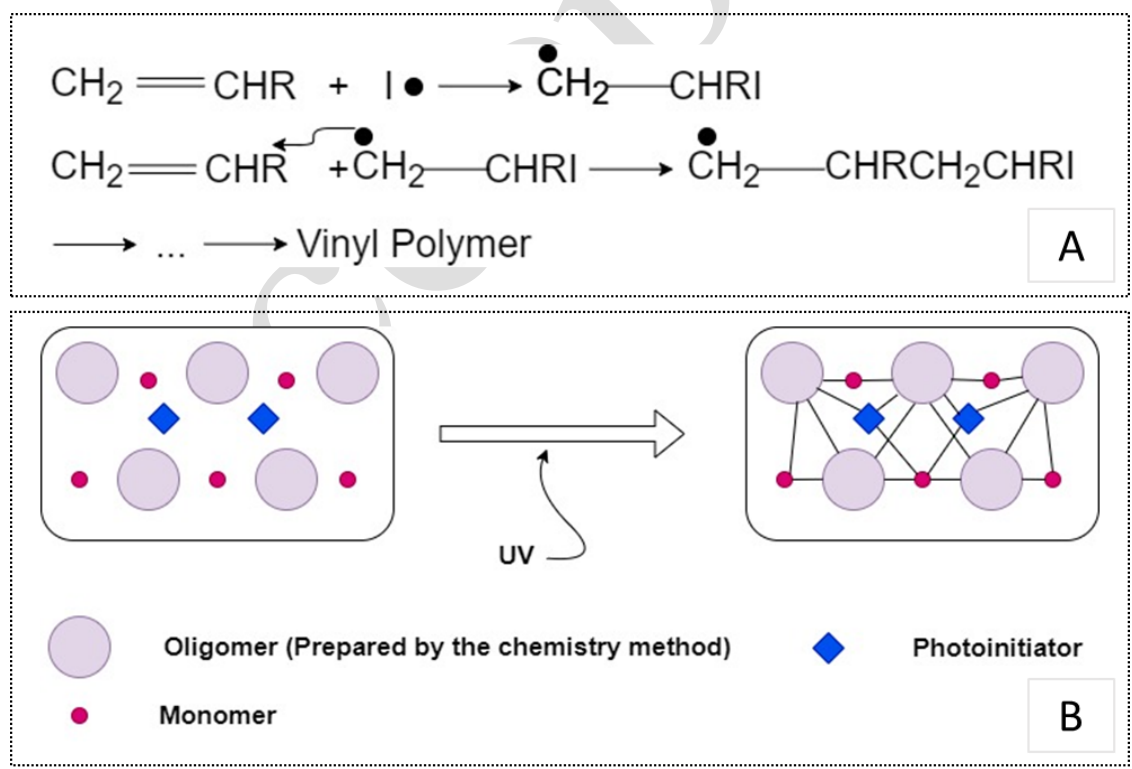

**Figure 9.** (A) General chemistry involved in the synthesis of vinyl polymers. (B) Photopolymerization of the monomers.

Poly(methyl methacrylate) (PMMA) is the most commonly used vinyl polymer in dental 3D printing. It is polymerized from methyl methacrylate monomer via FRP or anionic polymerization using 2,2´-azo-bis-isobutyronitrile or n-butyllithium as initiators, respectively. PMMA is the most favorable material for printing denture base materials owing to the ease of its processing, low cost, lightweight, stability in the oral environment, and esthetic properties. However, PMMA has poor surface properties and weak mechanical properties; the latter may be circumvented by using additives such as polyetheretherketone, $SiO_2$, and $Al_2O_3$ [3,77,78]. Furthermore, the addition of titanium dioxide yields antimicrobial property.

*3.1.2 Styrene polymers*

There are two styrene polymers commonly utilized in dental 3D printing: polystyrene (PS) and acrylonitrile-butadiene-styrene (ABS). PS is an aromatic hydrocarbon polymerized from styrene monomers via FRP, using benzoyl peroxide as initator (**Figure 10**). It is structurally amorphous with high transparency and a smooth surface. Mechanical properties and ease of fabrictaion make it a good candidate for dental applications [79,80]. High impact PS was used to prepare 3D printed objects using fused filament fabrication, with improvement of the properties of the resulting product [81]. Photopolymerization strategies are similar to those discussed in the previous session.

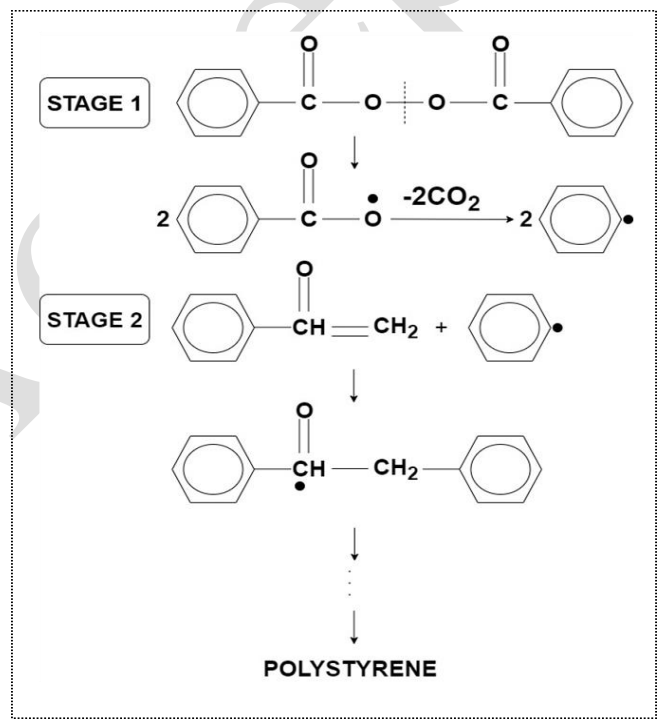

**Figure 10**. Free radical polymerization of polystyrene from styrene monomer and benzoyl peroxide as initiator.

Another material used in 3D printing is ABS, a thermoplastic polymer that inherits its superior properties from its monomers: acrylonitrile, butadiene and styrene [82,83]. Each of the monomers contributes to the quality of ABS. Therefore, ABS owes its heat tolerance, high impact strength, and rigidity to acrylonitrile, butadiene, and styrene respectively. Different initiators have been used for the synthesis of ABS via FRP: t-butyl or cumene hydroperoxide/sodium formaldehyde sulfoxylate dihydrate/EDTA-chelated $Fe^{2+}$ redox, ammonium persulfate, ammonium persulfate/sodium bisulfite redox initiator and oil-soluble 2.2'-azobis(2,4-dimethylvaleronitrile). FDM and SLS are usually employed for 3D printing of ABS [84–86].

*3.1.3 Polyesters*

Polyester refers to a group of thermoplastic polymers that contain ester functional groups in the main chain. They are polymerized via polycondensation by the removal of water molecules (**Figure 11**). The three most popular polyesters are polycarbonate (PC), polycaprolactone (PCL) and polylactic acid (PLA) [87,88].

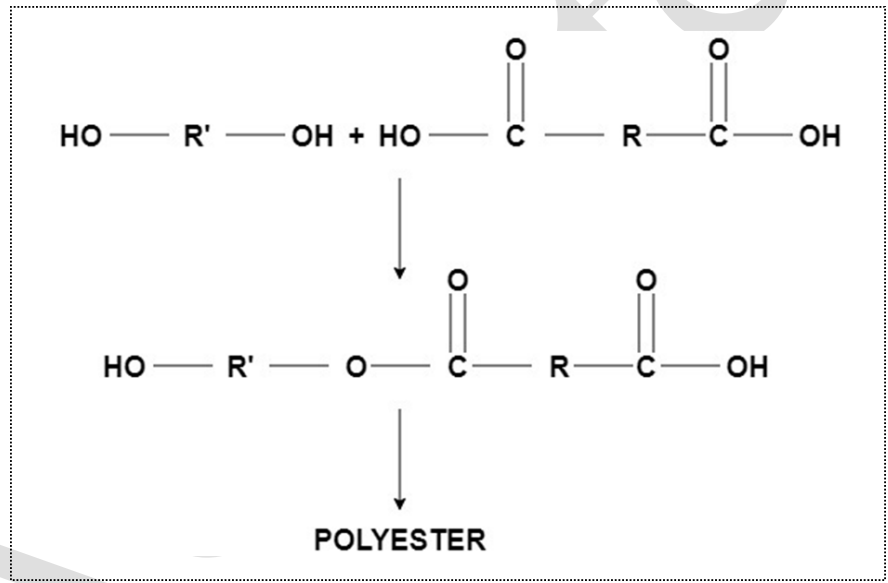

**Figure 11.** Synthesis of polyesters by polycondensation.

PCs are polymerized via polycondensation reaction between bisphenol A and carbonyl chloride/diphenyl carbonate from the main chain. They are mechanically robust, amorphous and transparent polymers. PCs are used extensively in dentistry to produce orthodontic brackets, denture bases and prefabricated provisional crowns. They have been used experimentally as a based composite printed by FDM [89]. Despite the advantageous properties of PC, there is a probability of bisphenol A release making it a potentially harmful substance [90].

PCL is synthesized via ring-opening polymerization of ε-caprolactone monomers in the presence of a catalyst such as stannous octanoate. PCL is a biodegradable and biocompatible polyester with high *in vivo* stability because of its hydrophobicity. In the field of 3D printing, PCL is of interest because of its low melting point (~63°C), making is useful for printing techniques such as FDM. Furthermore, 3D-printed PCL is used in bone tissue regeneration such as alveolar bone augmentation [91].

PLA is one of the most extensively used polymers in human body-related applications [92,93]. It is synthesized from lactic acid via polycondensation, and in some cases via ring-opening polymerization. The latter results in lower molecular weight and more brittle PLA. PLA is highly biocompatible and possesses tunable physicochemical properties. The excellent processability of PLA enables it to be used in different 3D printing methods and for various applications as FDM printing of drill guides for surgical insertion of dental implants, [94] and provisional restorations for protecting teeth after crown preparations [95].

### 3.2 Metal-based materials

Metals are vastly employed in fabricating biocompatible devices, especially when resistance to corrosion, and wear is required [96,97]. Mechanical properties and biocompatibility are also important factors for choosing metal for biomedical applications. Biocompatibility of a metallic material enables the material to perform its desired function without eliciting undesirable local or systemic effects on the surrounding tissues. Despite many types of metallic materials available, only a few of them are biologically compatible with the human body and can be used for long-term applications [98].

*3.2.1 Stainless steel alloys*

The earliest metal implant, introduced in the early 1990s, was made of vanadium steel [99]. However, it was associated with premature loss of implant function, patient dissatisfaction and a need for revision surgery because of mechanical and corrosion failure. As a consequence, stainless steel alloys with more advanced characteristics were introduced [100,101]. 302 stainless steel alloy was initially developed as an alternative to vanadium steel [102]. It was subsequently replaced by two other stainless steel alloys, 316 and 316 low carbon content (L). These alloys contain molybdenum and at least 11% chromium yielding corrosion resistance [103]. Formation of a protective Cr-based oxide film on the alloy surface occurs spontaneously in the body environment. This passive surface layer improves corrosion resistance and wear resistance of the fabricated implant [99]. Moreover, a maximum amount of 17-20% nickel helps to stabilize the austenitic phase at room temperature and increases corrosion resistance [103]. Nevertheless, the presence of a large amount of Cr and Ni in these stainless steel alloys can provoke an allergic reaction in patients, which require revision surgery for their retrieval [104]. These stainless steel alloys are also not suitable for manufacturing permanent implants [105]. Their low mechanical properties render them susceptible to stress, corrosion, cracking and pitting corrosion, with formation of deep pits on the metal surface. Implant failure initiates from these weakened spots along with the passive surface layer, which propagates through the bulk material [106].

*3.2.2 Titanium and its alloys*

The use of stainless steel as implants was eventually replaced by titanium and its alloys due to their improved corrosion resistance, biocompatibility, strength, and lightweight [107,108]. Commercially pure titanium (CP-Ti) and Ti-6Al-4V are the most biocompatible titanium alloys. They are the preferred materials for manufacturing orthopedic implants [109,110]. In addition, the formation of a thin, stable oxide film on the surface of titanium and its alloys contribute to their excellent corrosion resistance [111]. Bulk titanium alloy has a much higher Young's modulus than a natural bone (110 vs. 10-30 GPa, respectively) [112]. making it inapplicable for orthopedic use, since materials with Young's moduli exceeding that of a natural bone cause stress-shielding and subsequently bone resorption, which results in loosening of the implant. Despite the advanced properties of Ti-6Al-4V it cannot be used without surface treatment and coating due to its poor wear resistance. There are also potential concerns with toxicity of the vanadium or aluminum present in the alloy [113]. The second generation of titanium alloys have since been introduced that have lower moduli of elasticity (e.g., Ti-12Mo-6Zr-2Fe, Ti-15Mo-5Zr-3Al, Ti-15Sn-4Nb-2Ta-0.2Pd and To-13Nb-13Zr).[114] However, these alloys are more embrittlement-susceptible due to the oxygen dissolution caused by the oxygen diffusion into titanium which is introduced during fabrication/heat treatment. In addition, their higher cost have led to their substitution with cobalt-based alloys [115].

*3.2.3 Cobalt-based alloys*

Cobalt-based alloys have attracted considerable interest as medical implant materials because of their superior corrosion and wear resistance, excellent mechanical properties, biocompatibility and low rigidity [116]. The first cobalt chromium (CoCr)-based implant was fabricated in 1930. More than half a century of research has led to refinement in the manufacturing of this alloy for biomedical use [117].

Four types of CoCr-based alloys are available for biomedical applications, including cast CoCrMo alloy (F75), wrought CoCrWNi alloy (F90), wrought CoNiCrMo alloy (F562) and wrought CoNiCrMoWFe alloy (F563) [118]. Among the CoCr alloys, cobalt chromium molybdenum (CoCrMo) alloys are of major interest due to their advanced corrosion and wear resistance, high tensile and yield strength, and biocompatibility [104]. In addition, integration of oxide-forming elements such as Cr, Mo and Co makes them thermodynamically favorable for the formation of a thin protective passive oxide film after exposure to body fluids. This protective oxide layer is mostly composed of $Cr_2O_3$ with minor amounts of Co- and Mo-oxides [119].

Cobalt is the major alloying element of CoCrMo alloys. Cobalt imparts an unstable face centered cubic (FCC) crystal structure with very low stacking fault energy to the alloy. At a temperature of 690 K, Co-based alloys experience a change in their crystal structure, from unstable FCC to a hexagonal close-packed (HCP) crystal structure. This transformation occurs at relatively slow cooling rates. Since the transition from FCC to HCP is kinetically arrested at room temperature, the main crystalline structure of Co-based alloys is unstable FCC structure [120]. Transformation from FCC to HCP crystalline structure only occurs at high temperature after long-term exposure and/or application of mechanical stress. Both methods provide atoms with enough time to rearrange themselves from an unstable to a metastable structure, which leads to significant improvements in yield strength and reduced fatigue damage under cyclic stresses [121,122]. These properties account for the wear resistance and corrosion resistance of cobalt-based alloys [122].

Chromium has a significant effect on the alloy's properties [123]. Presence of ~28% chromium is the main reason for their corrosion resistance, which is characterized by the formation of a protective oxide film on the surface of the alloy segregating the surface from aggressive agents. Chromium is also the predominant carbide former. The embedded carbide in the alloys matrix serves as structural hardening sites that imparts additional strength to the matrix. Chromium-rich carbide, especially $M_{23}C_6$, is the most common carbide in CoCrMo alloys [124].

Molybdenum is another important constituent of CoCrMo alloys. It typically comprises 5-7 wt% of the alloys [125]. Molybdenum enables solid solution strengthening of the alloy due to its large atomic size. It serves as structural blocking sites for dislocation flow when present as solute atoms. In addition, formation of an Mo-oxide layer on the alloy surface imparts corrosion resistance to the alloy. If Mo is higher than 5 wt%, $M_6C$ carbide may also precipitate during alloy fabrication, which adds strength to the matrix [126].

Carbon is also an important constituent in the chemical composition of the alloy. Carbon is a FCC stabilizer which strengthens the alloy by forming carbide within the matrix. Based on the carbon content in the alloy, CoCrMo alloys are categorized into 'low carbon alloy' (0.06 wt%) and 'high carbon alloy' (0.15-0.25 wt%). Increase in carbon content results in improvement in corrosion, wear and tribocorrosion resistance of the CoCrMo alloy [127]. Despite having exceptional corrosion and wear resistance and excellent biocompatibility, CoCrMo alloys have poor frictional properties and may cause an allergic reaction due to their nickel content [128].

Metal-based AM techniques such as SLS and SLM are used extensively for fabricating 3D structures from a starting mixture of metal alloy powders characteristics of which are critical for reproducibility of the printed metallic parts. Metal powders used in 3D printing need to be spherical and have a stringent particle size distribution to achieve good packing behavior. This enables fabrication of dense and complex metallic parts with predictable mechanical properties. With the use of metal-based AM techniques, it is possible to fabricate fully-functional components using laser sintering in a layer-by-layer process.

The properties of AM alloys used for dental application have been investigated in different studies. For example, microstructure and hardness of CoCrMo dental alloys manufactured by casting, milling and SLS have been evaluated [129–131]. The highest hardness value was achieved by SLS (371 ± 10 HV), followed by casting (320 ± 12 HV) and milling (297 ± 5 HV). In another study, SLS-fabricated CoCrMo alloys had better mechanical properties and less dissolution of metal ions than cast alloys [132]. The total Co ions released into solution was lower in alloys produced by SLS [132]. Both groups of authors suggested SLS as a better fabrication method for dental devices than the conventional casting technique.

**3.3 Ceramics**

The use of ceramics has become very common in dental clinical practice as well as in research. Ceramics are highly biocompatible and possess strong mechanical properties, excellent wear resistance and superior esthetics compared with titanium or CoCr alloys [133]. These advantages have been harnessed in the development of 3D printed ceramic systems [10,134]. The following section discusses zirconia and alumina as ceramic materials for dental application.

*3.3.1 Zirconia*

The use of zirconia for fabricating metal-free ceramic dental implants began with the introduction of the Z-System in 2004 [135]. These implants osseointegrate well with hard tissues and are highly biocompatible with soft oral tissues. Different methods of treating the implant surface were employed to enhance osteoblast adhesion, differentiation and to improve bone-implant osseointegration [136]. Three major surface treatment methods have been employed: surface roughening, surface coating with active compositions to convert a bio-inert surface into a bioactive surface, and reducing surface contaminant to enhance surface hydrophilicity [137]. These methods include surface treatment with oxygen plasma,[138] acid-etching,[139] ultraviolet irradiation or hydrogen peroxide treatmet [140]. A previous study indicated that zirconia implants osseointegrate at least as well as titanium implants [141]. The strength of zirconia implants lies in their ability to accommodate peri-impant soft tissues well. *In vitro* data reported that zirconia decreases inflammatory response, plaque accumulation, reduce bacterial population and modify fibroblast adhesion and proliferation [142,143]. As an example, yttria-stabilized zirconia (3Y-TZP) powder has been blended with acrylates and methacrylates, photoinitiator and dispersing agent to create a homogenous hybrid sol. This "ink" can be photopolymerized by a 3D printer with light-curing capability. SL was used for printing dental bridges (**Figure 12**) [133].

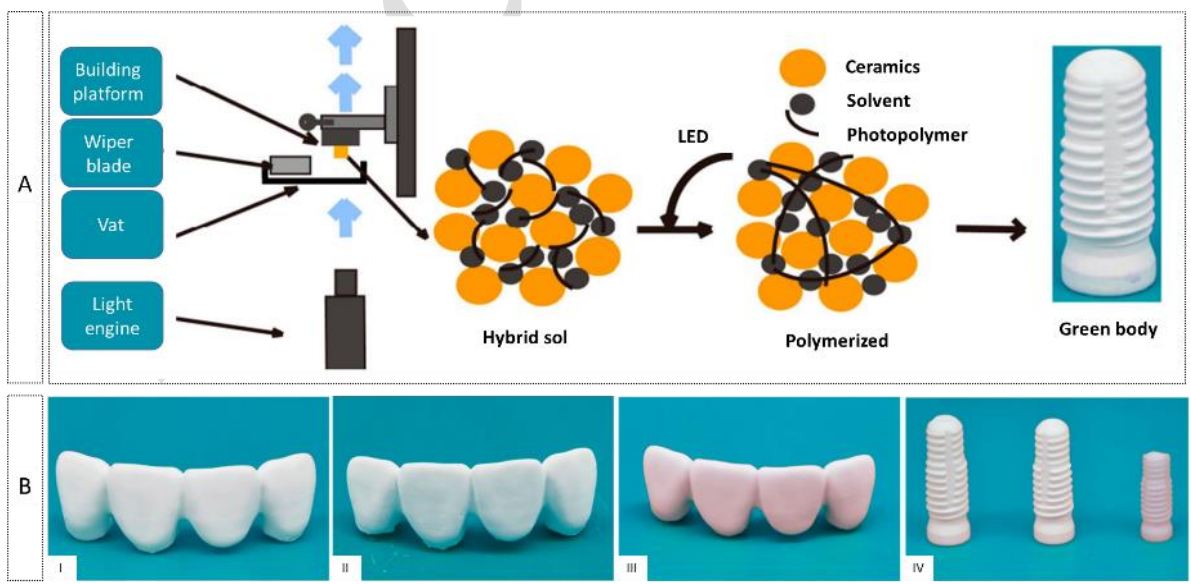

**Figure 12.** The use of ceramic (zirconia) for 3D printing of dental bridges. (A) An illustration of the ceramic stereolithography process. (B) Additive manufacturing of zirconia dental bridges after (I) printing, (II) debinding and (III) sintering. (IV) The corresponding change in size of zirconia-based dental implants after each step. Reprinted with permission from [133].

*3.3.2 Alumina*

Aluminum oxide or alumina ($Al_2O_3$) is a ceramic material produced from alumina trihydrate. It is available in different forms (e.g., α, χ, η, δ, κ, θ, γ and ρ). All these forms exist during heat treatment of aluminum hydroxide [144]. The most thermodynamically stable form is α-aluminum oxide. This material is used for such dentistry applications as ceramic abutments, endodontic posts, orthodontic brackets, dental implants, and crowns [134]. The high purity of alumina (~99.99% pure) makes it a suitable candidate for replacing metal alloys in some dental applications [33]. Alumina is wear-resistant and biocompatible but is less compact and has lower flexural strength compared with zirconia [145]. To improve the mechanical properties of alumina, it is possible to infiltrate zirconia into the alumina matrix to produce a mineral composite [146]. The toughness and fracture resistance of alumina may be altered by controlling the grain size and porosity, the number of sinter cycles, sintering temperature and time, heating/cooling rates, and the addition of stabilizer such as zirconium oxide, magnesium oxide or chromium oxide [147,148]. An alumina ceramic with submicron grain-size has been prepared for dental applications, with improvements in mechanical properties and optical quality. Studies on polycrystalline alumina ceramics indicate that alumina with grain size less than 1 µm and a porosity level of less than 0.7 percent has the same degree of translucency as commercial high-translucency porcelain [149].

**3.4 Responsive materials for 4D printing**

Four-dimensional printing adds the time dimension to 3D printing by producing pieces made of smart materials. 4D printing indeed combines 3D printing with the use of responsive materials that change their properties or shape in response to mechanical, chemical, thermal or electrical stimuli. Smart materials may be divided into shape-memory materials and shape-changing materials [11]. Shape memory materials may be derived from alloys, polymers or ceramics. Each material has advantages and disadvantages depending on the applications of the material. Shape memory polymers have applications in dentistry including orthodontics, endodontics, prosthodontics, oral surgery and implantology [150–152]. Through molecular view, shape memory polymers are elastic matrixes enjoying some reversible functional covalent cross-linking. This kind of polymers consists of functional groups that undergo some changes toward a temporary shape under a stimulus. They can subsequently reverse that change to the previous permanent shape once the stimulus is canceled; or they undergo cleaving the covalent bond induced by the stimuli between the functional groups throughout the matrix [153]. To prepare a 3D printed object that responds to external or internal stimuli, sensitive materials should be employed in these systems (**Figure 13**). Some examples are presented n the following paragraphs.

Magnetically-sensitive materials are generally constructed by embedding magnetic nanocompounds (e.g., Fe, Co, Ni, $Fe_2O_3$, $Fe_3O_4$, $CuFe_2O_4$, $ZnFe_2O_4$, $NiFe_2O_4$) into the raw material. These nanomaterials are utilized *in-situ* (embedded) or decorated/mixed with the pre-defined materials for printing [154,155]. Electrically-responsive materials are manufactured by incorporating inherently-conductive polymers (e.g., polyacetylene, polypyrrole, polyaniline, polythiophene) as well as their derivatives or copolymers into the material. Conductive carbon-based nanostructures and metal nanoparticles may also be used for this purpose [156,157].

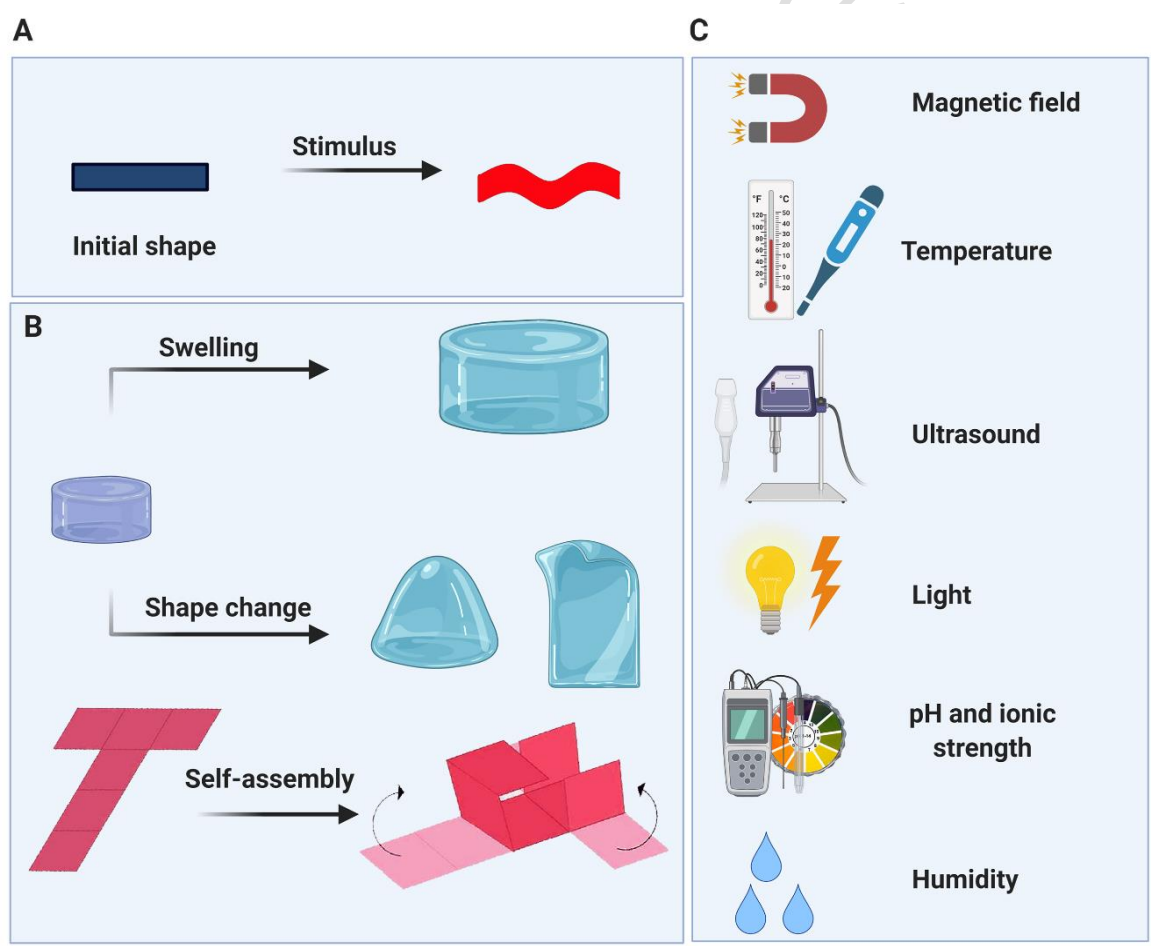

**Figure 13** (A) Schematic of the deformation of a 4D-printed object in response to a stimulus. (B) Types of responses that may occur in 4D-printed dynamic materials. (C) Internal or external stimulus that may be utilized for triggering a 4D response.

Light-responsive materials are fabricated from light-conducting materials or materials comprising azobenzene, stilbene, spiropyran, fulgide or diarylethene, as well as photosensitive-metal nanostructures (e.g., TiO$_2$, Pt and Au) [158,159]. Ultrasound-sensitive materials comprise polymers the bonds of which are cleaved in the presence of high ultrasound intensity. These polymers are classified into biodegradable (e.g., polyglycolides and polylactides) and non-biodegradable polymers (e.g., poly(ethylene-co-vinyl acetate) and poly(lactide-co-glycolide)) [160,161].

Thermo-responsive materials (e.g., poly(*N*-isopropyl acrylamide), poly(*N*-vinylcaprolactam), gelatin, collagen, soybean oil epoxidized acrylate, pluronic and poly(ether urethane) are frequently employed materials in 4D printing platforms [162–164]. pH-sensitive materials swell or collapse depending on the surrounding pH due to the presence of functional groups such as hydroxyl (−OH), carboxylic (−COOH), sulfonic acid (−SO$_3$H) and amine (−NH$_2$) groups in the polymer chain [165,166]. Moisture-sensitive materials (e.g., hydrogels) are similar and have similar hydrophilic functional group as the pH-responsive systems. Their hydrophilicity causes the systems to swell to a volume greater than the original volume [166,167].

There is another class of materials that responds to a biological stimulus (e.g., glucose, enzymes). For example, the level of blood glucose may be controlled by using smart materials that release insulin in response to the blood sugar level [11].

The characteristics of smart materials include self-sensing, self-responsiveness, shape memory, self-repair, self-adaptability and multi-functionality. Smart materials fall into two major categories, shape-memory materials which recover their original shape following a stimulus and shape-changing materials which maintain their original shape and undergo morphology change in response to a stimulus [164,168].

Experimental dental 4D printing began in 2012. Although the technology is not yet commercially available, it represents a quantum leap in AM capability. For example, 4D printed dental implants possess the capability to alter their shape in response to changes in oral temperature and humidity. The properties that these implants possess are comparable to those exhibited by natural teeth [150,169].

**3.5 Materials Safty**

Ceramic materials used in dentistry are reported to be relatively safe and free of toxicity for dental pulp or periodontal tissue surrounding the material [170–172]. Corrosion is the main complication followed by metal devices in the oral cavity. Nickel and chrome are the most substances found in the oral cavity, due to placement of orthodontic appliances, metal crowns, metal bridge or prosthesis devices. Several other consequences might be followed with application of metal such as hypersensitivity, cytotoxicity, or genotoxicity. Exposure of oral mucosa to metal leads to inflammation, pain, redness or at a deeper level causes cellular metabolism disturbance. Depending on the degree of corrosion that metal undergoes, its carcinogenesis ability varies. The symptoms can appear at distance from metal alloy [173–175].

Polymer materials have various applications in dentistry, however, their main form of application in 3D printers are light curable resins. The uncured resins including monomers, macromonomers, and initiators are not biocompatible materials. After photocuring, the degree of conversion is relatively high which means the amount of unreacted monomers is decreased; however, the residual monomers may lead to irritation of mucosa, inflammation, ulceration and edema. The side effects of residual monomers are not only limited in local tissue and affect systemic circulation in case of blood stream entry [176]. This situation may be better for the 3D printers (e.g., FDM method) that do not employ resin. At least, for the filaments used for FDM approach, there are no cytotoxic monomers but the type of polymer/composite employed for filament should be taken into consideration. However, for a specific 3D printer, most providers sell a different range of resins from biocompatible (for medical applications) to non-biocompatible compounds (for other applications). Accordingly, based on the targeted applications, the desired materials can be selected.

## 4. Applications

Over the last decade, 3D printing technology is increasingly being utilized in dental education and patient treatment. Examples include the manufacture of drill guides for prosthodontics, orthodontics, endodontics and oral maxillofacial surgery. New technologies have been developed for the production of dentures and implants, and the fabrication of bridges, crowns, aligners and orthodontic brackets [34,177]. The recent advances in the applications of 3D printing in dentistry will be highlighted below. **Table 2** represents the applications of 3D printing in dentistry.

**Table 2.** Recent applications of 3D printing in dentistry.

| Method | Schematic illustration | Applications | References |
|---|---|---|---|
| Stereolithography | 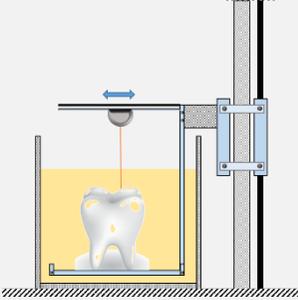 | Temporary and permanent crowns<br>Temporary bridges<br>Temporary restorations<br>Surgical guides and templates<br>Dental replica models | [3,27]<br>[178–180]<br>[181,182]<br>[22,26,183] |
| Fused deposition modeling | 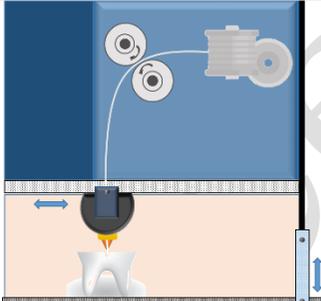 | Denture flasks<br>Bites<br>Mouth guards<br>Oral drug delivery device | [133]<br>[14]<br><br>[184] |
| Selective laser sintering | 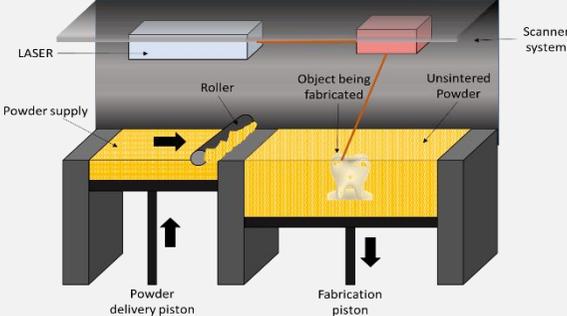 | Custom-made dental implants<br>Partial dentures | [185–187] |

| Photopolymer Jetting | 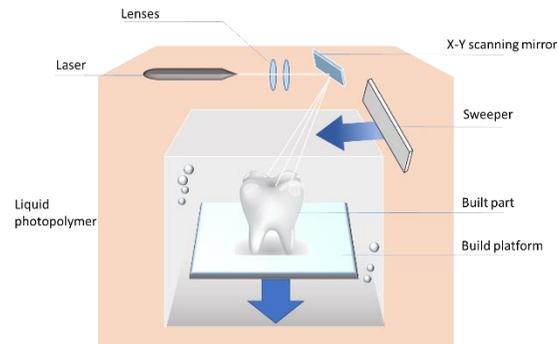 | Temporary crowns<br>Dental replica models | [27,188–190] |
|---|---|---|---|
| Powder Binder Printer | 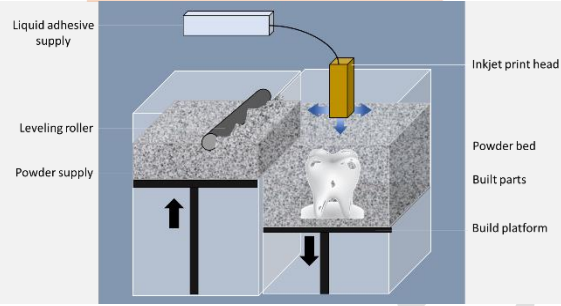 | Study casts | [18,191] |
| Digital Light Processing | 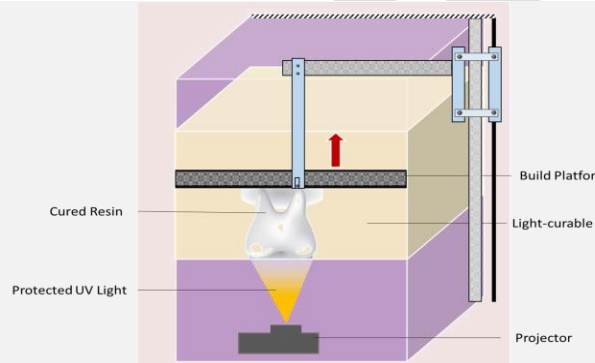 | Temporary and permanent crowns<br>Removable dental prosthesis | [192]<br>[193] |

## 4.1. Dentures

Tooth loss occurs due to trauma, fracture, periodontal disease and caries [194,195]. Different methods have been used to replace missing teeth, such as complete or partial removable prostheses, teeth-supported bridges and implant-supported crowns. In the past, construction of a prosthetic device commences with the taking of an alginate or polyvinylsiloxane impression. A model is subsequently constructed for complex laboratory work performed by a skilled dental technician. The advent CAD/compute-aided manufacturing (CAD/CAM) systems has completely revolutionalized how dental prostheses are constructed. There are three phases in the construction scheme. The first phase involves acquisition of virtual impressions using intraoral and extraoral scanners and registering the patient's occlusion. The second phase consists of prosthesis design using computer softwares. These procedures may be performed in the the laboratory or a dental clinic. This enables the practitioner to have better visualization of the final product. The third phase involves prosthesis construction, which may be performed using additive or subtractive manufacturing techniques [196–198].

Denture design and manufacturing now involves the use of 3D printing technology in combination with digital modeling and computational optimization. Both denture bases and denture teeth are now printed by 3D printing technology using methacrylate-based photopolymerizable resins [199,200]. Denture based and artificial teeth are printed separately and are subsequently attached together using a photo-curable bonding agent. Unlike conventional denture processing technique which involves compression molding and hot water processing, 3D printed dentures enables more efficacious clinical modification, which reduces patient discomfort as well as long-term residual bone resorption (**Figure 14**) [34,177,199–201].

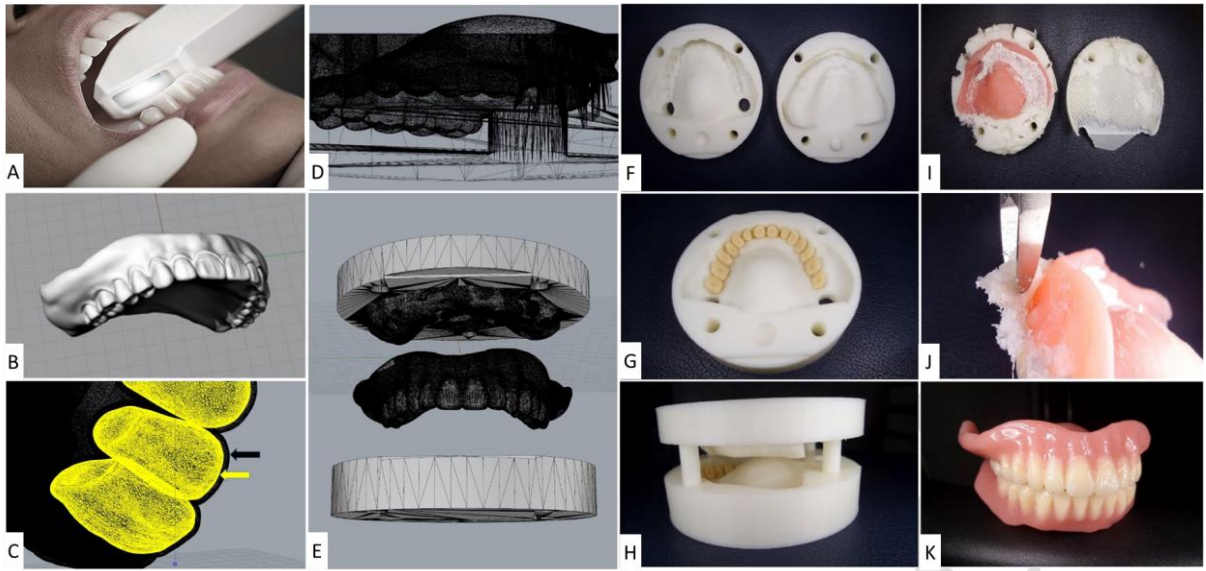

**Figure 14.** (A) Intraoral scanning for taking virtual impression of a patient's pre-existing teeth. (B) 3D illustration of denture file from a design software. (C) A real tooth size configuration used for reconstruction of the denture. (D) Digital sculpting of denture teeth secured to a denture flask. Pink denture base acrylic resin is to be injected through the injection port. (E) Separate illustration of the denture and flask from the analyzing software. (F) 3D–printed flask. (G) Placement of the teeth into the flask. The teeth were fabricated also by 3D printing. (H) Combination of upper and lower denture flasks produced by 3D printing. (I) Separate printed upper and lower flasks showing packing of acryic resin. (J) Detachment of the flask and denture parts. (K) Polished, completed 3D printed denture. Reprinted with permission from [184,202].

## 4.2. Crowns and bridges

3D printing of dental crowns and bridges is one of the most attractive applications of 3D printing technology in dentistry. Crowns may be placed on teeth for protecting the underlying tooth structure, or on dental implants to make them functional and tooth-like. One of the most important uses of crowns is to serve as an anchor (i.e. tooth abutment) for replacing a missing tooth. Recently, 3D printing technology has been utilized to fabricate crowns and bridges. In addition, low-cost 3D printers have been used for fabricating more precise provisional crown and bridge restorations using SLA printing techqniue [203]. SLA offers great efficiency and high level of accuracy [204]. By using SLA, an object can be produced at a resolution down to 0.05 mm, making this method a superior technique in accuracy, even compared to some new techniques such as digital light projection (DLP) [205]. However, the process of photopolymeter curing in DLP printer is fairly faster than SLA, since SLA uses laser light photopolymer sources that process layer by layer [206].

**4.3. Orthodontic appliances and orthotics**

3D printing provides exciting opportunities for fabricating tooth-fitting orthodontic devices such as Invisalign® aligners [207]. Aligners are popular alternatives to braces for treatment of mild malocclusion because they are transparent and can be removed for eating, drinking and tooth brushing. In 3D printing of clear aligners, the orthodontist takes a virtual impression of the patient's upper or lower arch using an intraoral scanner. Stereolithography or FDM is then used to print the model for thermoforming the aligner. 3D printing reduces production time without altering the quality of the aligners. This acceleration is attributed to the reduction in manufacturing steps (**Figure 15**).

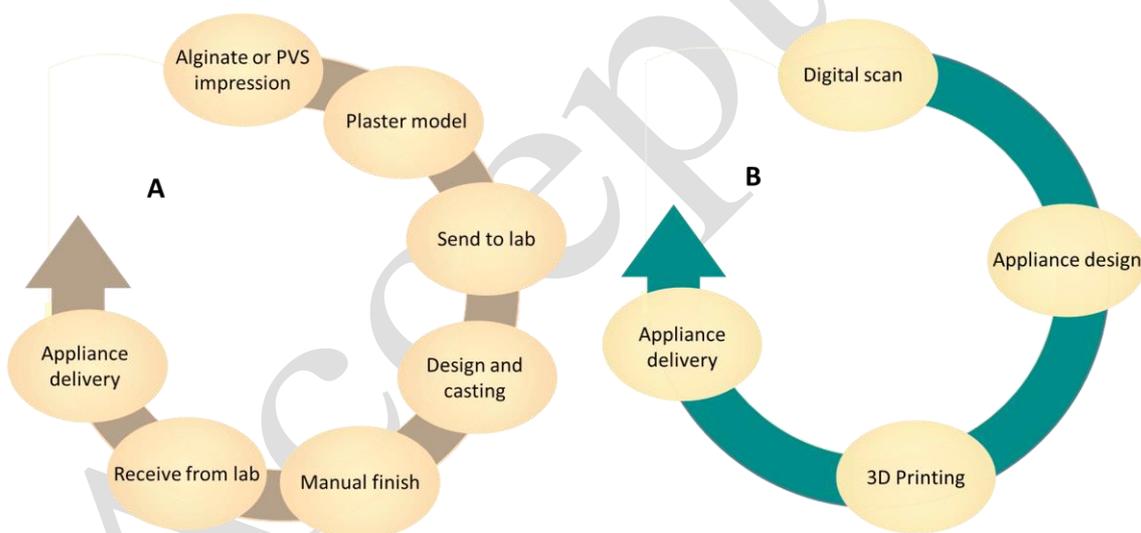

**Figure 15.** Designin an aligner. (A) Steps involved in conventional processing (B) Workflow in 3D printing.

Orthotics are used for treatment of malocclusion and relief of temporomanidular joint symptoms. These appliances can also be fabricated by the use of intraoral scanners and 3D printers. Nanofillers used for reinforcing these appliances, such as silica and hydroxyapatite, do not bind to the resin during SLA. In light of this, the surface of HAp was decorated with methacrylate functionality to enhance adhesion of the nanofillers to the polymer matrix (**Figure 16**) [14]. Quaternary ammonium salts have also been functionalized on the nanofillers to impart antibacterial and antifungal activities. The resulting orthotics exhibited enhanced mechanical properties and possessed antimicrobial activity [14].

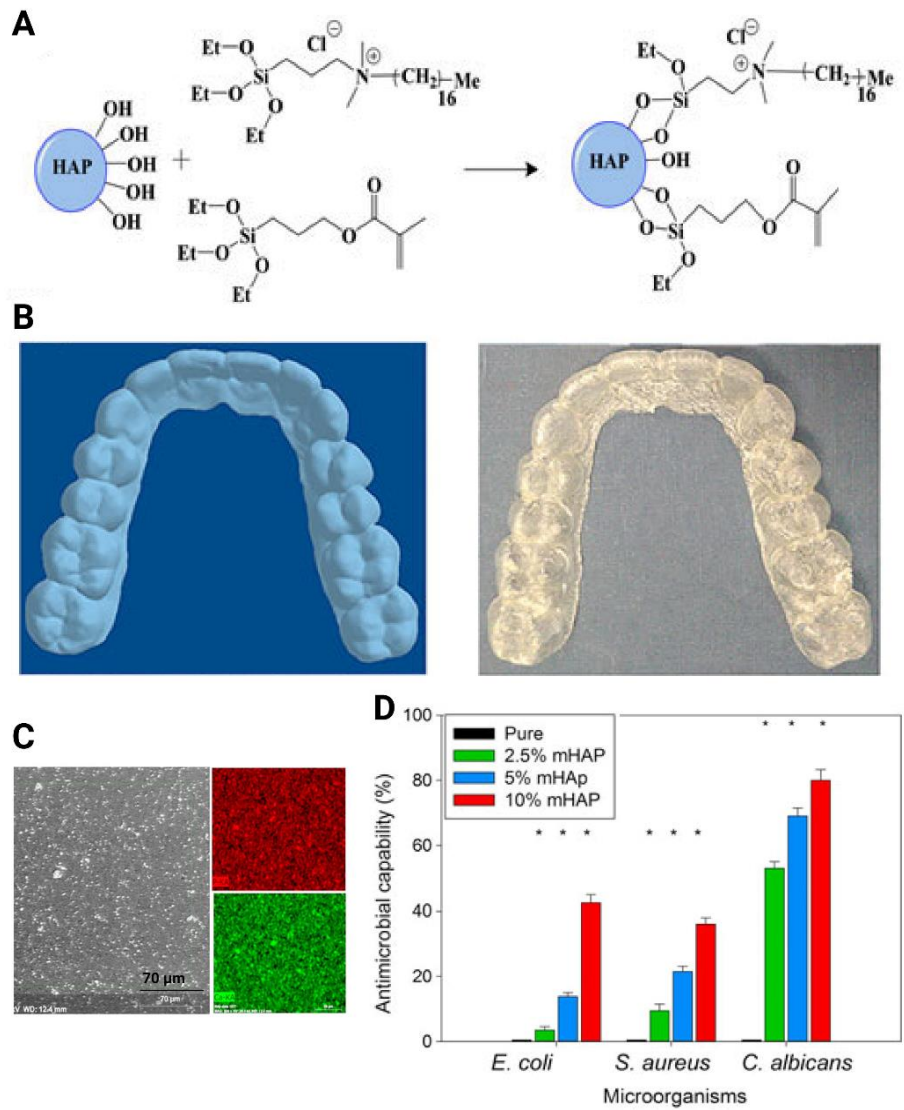

**Figure 16**. (A) Surface functionalization of hydroxyapatite. (B) 3D CAD drawing of an orthodontic aligner (left panel) and the photograph of the corresponding aligner fabricated by SLA (right panel). (C) Scanning electron microscopy image of the material used for construction of a nanofillers-reinforced orthotic (left panel). Energy dispersive X-Ray (EDX) mappings of P (red) and Ca (green). (D) Antibacterial and antifungal properties of the samples containing different ratios of the nanofiller. mHAP: modified HAP. Reprinted with permission from [14].

### 4.4. Oral and maxillofacial surgery

The advent of cone beam computerized tomography (CBCT) has expedited the development of 3D printing in dentistry. This has resulted in more accurate diagnosis and treatment planning, as well as guidance during surgery, thereby reducing post-operative complications. The types of surgery that benefit from the use of 3D printed devices include maxillofacial surgeries, implant placement in difficult or ethetics-sensitive areas, sinus lift and orthognathic surgeries.

Maxillofacial defects caused by trauma or cancer require extensive removal of the defect as well as the surrounding tissues. Depending on the size of the defect, the severity of the trauma and the patient's economical condition, oral, nasal and orbital maxillofacial prostheses are often required for post-surgical rehabilitation. The recent use of scanners for scanning the area of involvement is a much more comfortable and precise means for fabricating the prostheses. Maxillofacial prosthesis printed using AM methods have been shown to have higher accuracy and better fit of the defect area [208,209].

Orthognathic surgeries are performed for patients with maxillary or mandibular deformity, severe malocclusion and those requiring facial esthetic modifications. Development of 3D printed models have improved patient communication as well as surgical outcome. Many of the traditional operations are now replaced with the help of digital techniques, such as printing of occlusal splints tht are more accurate and provide more accessible jaw positions during surgery (**Figure 17**). The use of 3D printed osteotomy or genioplasty surgical guides reduces operation time and minimizes injuries to nerves such as the inferior alveolar nerve[210].

The use of 3D printed surgical guides for implant placement in the edentulous area improves hard and soft tissue healing aound the implant. Surgical guides are used in challenging locations to improve implant placement accuracy. The surgical guide must have adequate hard or sift tissue support to increase stability and balance during implant placement. Support is gained by methods such as tooth-supported guides, mucosa-supported guides, bone-supported guides and pin -supported guides. These guides are selected based on the number of remaning teeth, the availability of hard and soft tissue support and the location of the implant site [211].

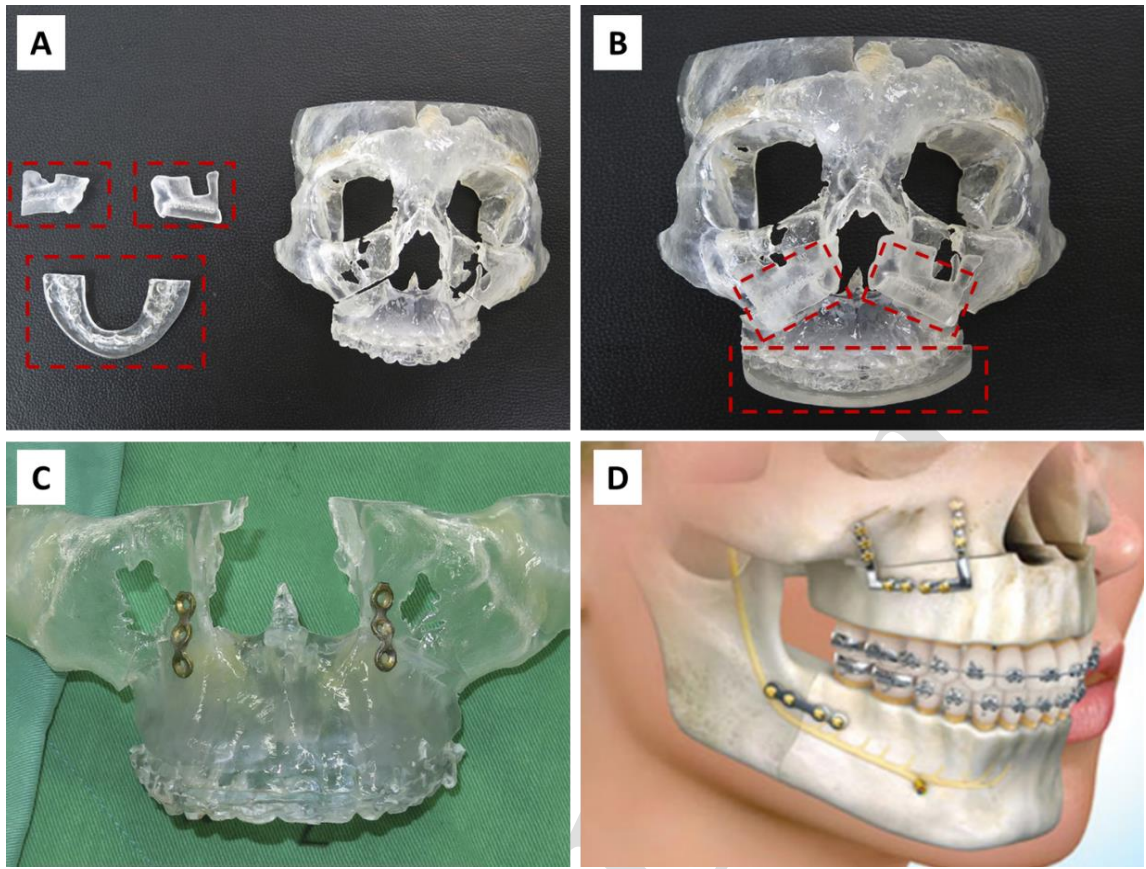

**Figure 17.** (A) 3D printing of a patient's jaw model after simulation for orthognathic surgery. (B) Positioning and fixing the printed guides and occlusal splint. (C) A 3D printed model of the final position of segment. (D) An illustration of the application of 3D printing in maxillofacial surgery. Reprinted with permission from [210].

### 4.4 3D printing of metals in dentistry

Accuracy and mechanical properties are the most important factors to be considered in dentistry. Compared to conventional casting and forging techniques, additive manufacturing of metal parts reduces manufacturing time and cost. More importantly, AM almost eliminates all human errors as well as possible defects in the resulting product. According to the definition of the American Society for Testing and Materials (ASTM), ASTM F2792, AM is "a process of joining materials to make objects from 3D model data, usually layer upon layer, as opposed to subtractive manufacturing methodologies"[212]. Selective laser sintering, selective laser melting (SLM) and electron-beam additive manufacturing (EBM) are the three most commonly used techniques for 3D printing of metal parts [213].

Dental metal alloys that utilize 3D printing technology include CoCr, Ti and tantalum powders (**Figure 18**) [214]. For example, dental prostheses have been fabricated with CoCr alloy powder using the SLM technique [215]. The potential of 3D printed metal parts in dentistry is enormous. Metallic structures that can be 3D-printed include removable partial denture, overdenture and fixed prosthetic frameworks, endodontic hand and rotary files and metal implants. Metal frameworks were prepared by SLM technique to evaluate the accuracy of removable partial denture frameworks. The authors showed that the frameworks manufactured by SLM fit better than those fabricated by the lost-wax technique and metal casting [216]. In another work, porous implants made of Ti and Ta were fabricated via SLM and tested for their fitting accuracy. The authors reported that 3-D printed Ta implants were better-fitting compared to conventional Ti implants manufactured by milling. By using 3D printing, the authors were able to optimize manufacturing cost and maximize reproducibility [217].

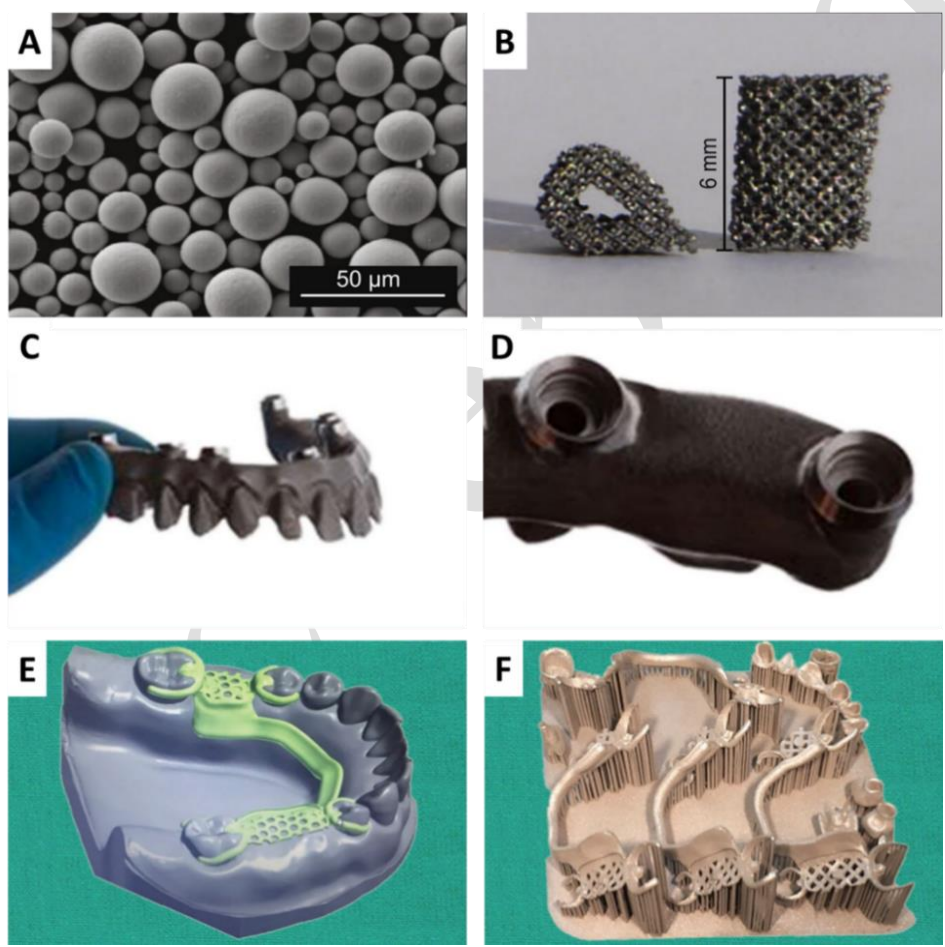

**Figure 18.** 3D printing of metal in dentistry. (A) SEM image of tantalum powder used for 3D printing of dental implants. (B) A 3D printed porous metal-based implant. (C, D) A 3D printed CoCr-based denture implant framework. (E) Designing a removable partial denture framework prior to 3D printing. (F) A 3D printed CoCr powder-based removable partial denture framework with accompanying support structures that have to be removed after printing. Reprinted with permission from [213,214,216].

### 4.4. Hard and soft tissue engineering

Bone augmentation is an important procedure in patients with hard tissue defects who are in need of implant placement. Insufficient bone results in lack of support and implant failure.

Hence, bone augmentation is necessary in selected cases prior to implant placement. The ideal bone graft to be placed is an autograft derived from the patient's own bone, which has both osteoinductive and osteoconductive properties. Other alternative bone graft materials include allografts, xenografts and alloplastics. A graft used for bone augmentation needs to have specific properties, including biocompatibility, same shape and design as the defect, does not induce an immune response, as well as porosities for tissues to access nutrients and communicate with immune cells (**Figure 19**) [218].

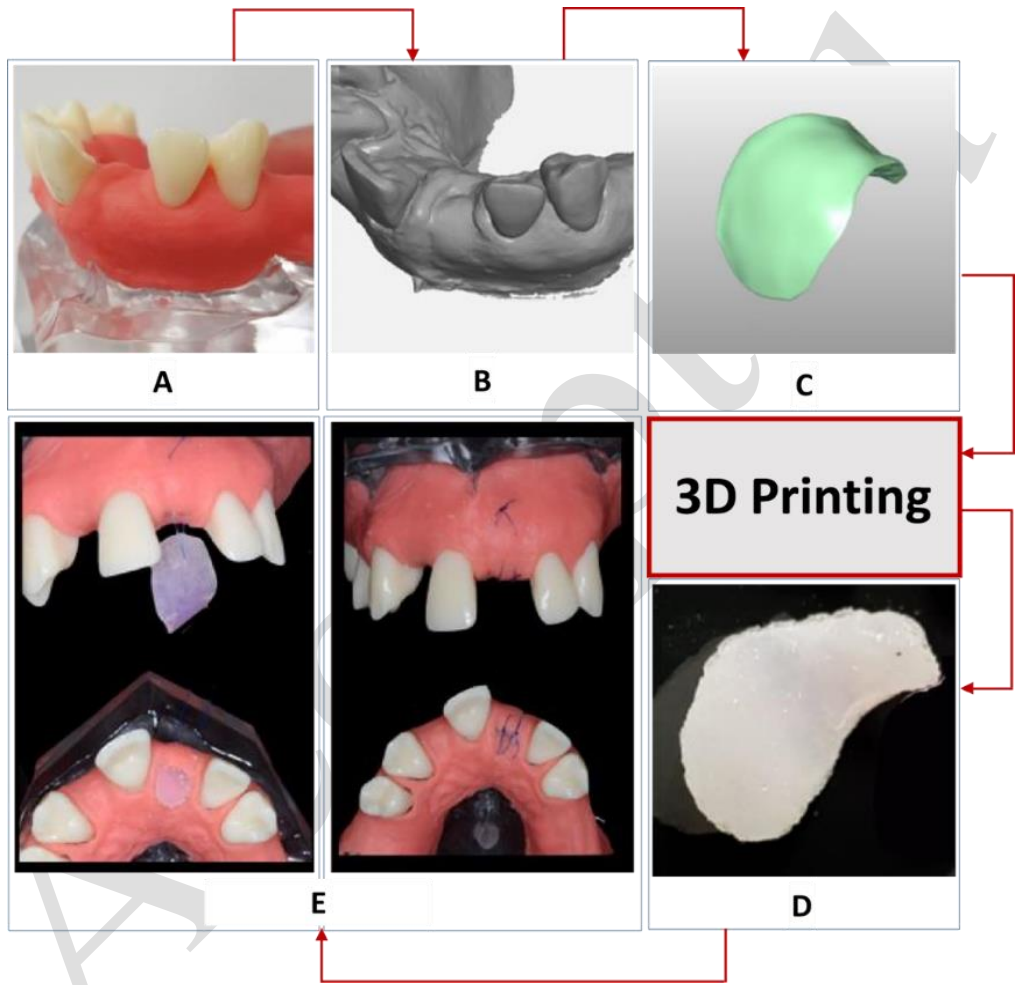

**Figure 19.** Tissue augmentation workflow. (A) The soft tissue defect. (B) Scanned soft tissue defect. (C) Designed soft tissue graft ready to be 3D printed. (D) 3D printed soft tissue graft. (E) Simulated surgical procedure. Reprinted and modified with permission from [219].

Many biomaterials are available as hard tissue engineering scaffolds, such as polymers, bioceramics and composites. To enhance the regeneration ability of the biomaterials, they should be able to be loaded with stem cells; growth factors released by the stem cells are critical for angiogenesis and osteogenesis. Scaffolds fabricated by 3D printers can be tuned by using inks containing stem cells. This method of printing enables the stem cells to be placed at precise locations. Moreover, precision 3D printing of complex scaffold surfaces enables more intimate contact with bone surfaces, expedites healing and yields better esthetic results compared with conventional scaffolds [220].

The periodontal ligament (PDL) absorbs masticatory stresses and provides the tooth with micro-movement during mastication. The PDL contains blood vessels and nerve fibers. Destruction of the PDL reduces tooth support, increases tooth movement and expedites microbial infection. Periodontal bone defects may be covered with a 3D printed scaffold in which mesenchymal stem cells are incorporated within the scaffold to enhance regeneration of the PDL.

Soft tissue regeneration surgery has been used to improve root coverage of a tooth with gingival recession or damaged soft tissue. Although the ideal graft is an autograft excised from the palate or tuberosity, this generates post-operation discomfort for the patient. Recently, 3D printed soft tissue grafts have been introduced for augmentation of keratinized tissue surrounding teeth with periodontal defects. These soft tissue grafts can be printed to cover large defects. Moreover, 3D printed soft tissue grafts permit more complicated designs and accuracy to cover the defect. They are also better received by patient because of reduced post-operation discomfort and expedited healing [219].

### 4.5. Mouthguards for drug delivery

Treating oral diseases by local release of a drug is common practice in oral medicine. 3D printing permits the design of unique drug delivery systems with exceptional precision, comprehensive three dimensional composition and manageable release patterns [83]. Chlorhexidine-coated mouthguards have been tested for suppression of oral bacteria in human subjects [221]. A 3D printed wearable oral delivery device in the form of a mouthguard has been fabricated via FDM to deliver a preloaded drug to the oral cavity (**Figure 20**) [184]. 3D printed oral drug delivery mouthguards has immense potential for drug delivery in personalized dental therapeutics.

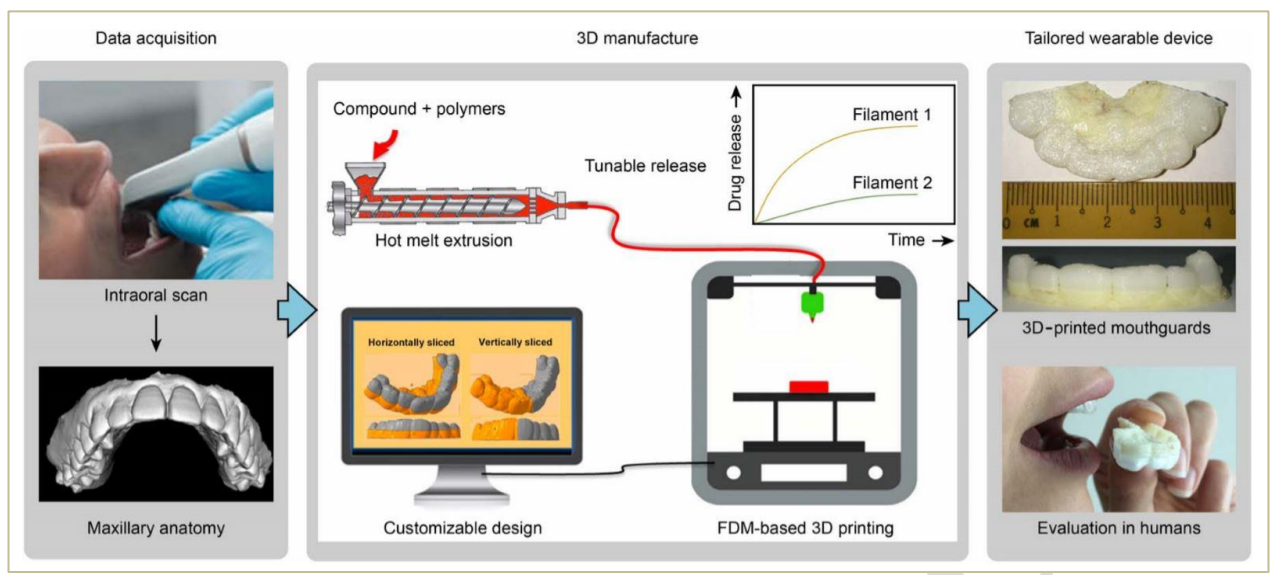

**Figure 20**. Application of 3D printing in manufacturing of oral drug delivery mouthguards. Reprinted with permission from [184].

## 4.6. 3D printing in endodontics

Three dimensional printing has been used to support nonsurgical and surgical root canal treatment. As an example, 3D printing may be used experimentally for regeneration of new teeth by stem cell delivery[219], or regeneration of the dental pulp by development of calcium phosphate cements [222]. In addition, 3D printing has been used for printing tooth models to simulate surgical procedures. For example, a tooth model was printed for root canal treatment of a 12 year-old boy with tooth anomalies. The authors designed and printed a custom-made guide for access into the canal spaces (**Figure 21**)[223]. The advantages of the 3D printing in endodontics include improving the accuracy of access, enhancing the skills of non-specialists and improved success with pulpal regeneration [224] In addition, 3D printed resin teeth are useful in educating students and general practitioners in all steps of the root canal therapy. The trainee can observe endodontic working length, root canal morphology and anatomy of the apical delta, using transparent resin teeth constructed usingh CBCT data derived from extracted natural teeth. During canal shaping, errors associated with the use of endodontic instruments, such as gauging of preparation walls and canal transportation may be readily observed. More importantly, 3D printed guides based on CBCT data are indispensable for finding highly calcified canals in nonsurgical root canal treatment and for apicoetomy of posterior teeth in surgical endodontics. The use of these guides has enabled more teeth to be saved by root canal treatment.

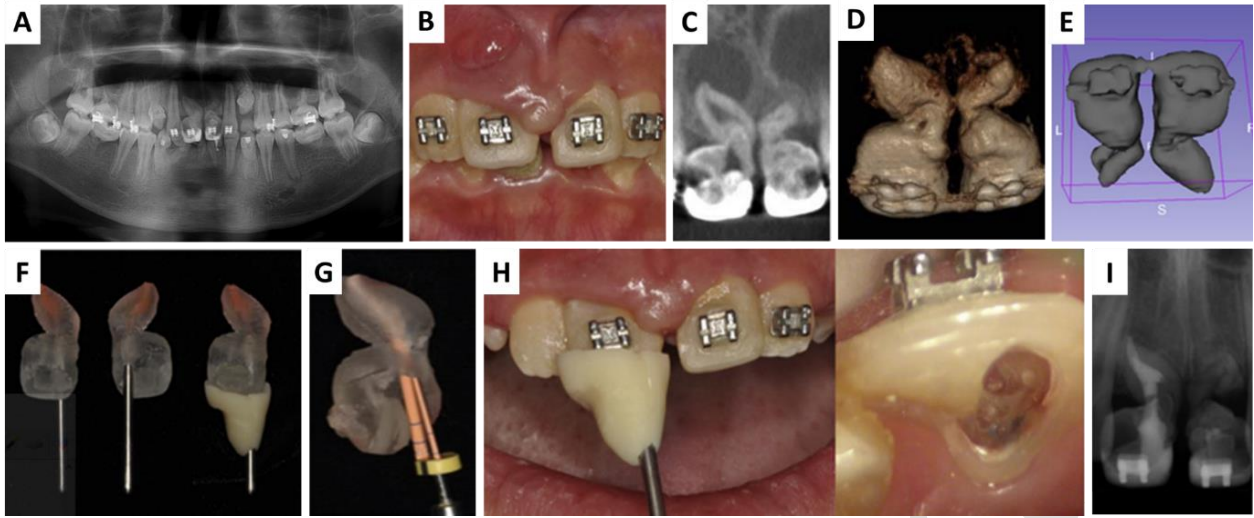

**Figure 21.** A 3D printed, custom-made jig for complicated root canal treatment. (A) Panoramic radiograph of the patient's teeth. (B) Buccal view. (C) Coronal view. (D) Digital image view. I STL file. (F) 3D printed jig with a hole that guides the insertion path and depth of a bur into the complicated root canal space. (G) Canal filing simulation. (H) Location of glide path into the root canal space using the printed jig; (I) Intraoral radiograph 3 months after canal filling. Reprinted with permission from [223].

### 4.7. 3D printing application in dental education

Three dimensional printing in dentistry has extended beyond clinical application and has been integrated into dental education [225]. 3D printed teeth serve as models for teaching dental anatomy and for the practice of a wide range of dental procedures such as caries excavation, pulp capping, core build-up, crown preparation as well as veneer preparation and dental bonding [226]. These models are used for training predoctoral students and have replaced the use of extracted teeth which are prone to contamination [227]. Apart from the anatomical accuracy of the printed replicas, fabrication of these resin replica teeth can be done by any institution that has access to CBCT and a 3D printer [228].

Three dimensionally-printed models are extensively used for pre-operative surgical simulations. The ease of fabricating these surgical models provides a readily available resource for oral and maxillofacial surgery [229]. It is beyond doubt that this versatile technology will keep on improving over the course of its application. The elastic properties of currently available materials and the geometric complexity of the printed models still require improvements for perfect mimicking of human tissues [230].

### 5. 4D printing in dentistry and maxillofacial surgery: Challenges and future perspectives

In a nutshell, 4D printing technology can give active and responsive functions to their counterpart 3D printed objects. It can be triggered by a response to different stimuli, for example, light, to change their structure. Fortunately, through using simulations and analytical calculations to theoretically model the structural changes are well predictable. In other words, we know what will happen for the material which at the mercy of any external stimulus. In dentistry, it is of the most interest to have such materials [231].

One of the most common materials used in orthodontics is nickel-titanium archwires, with many advantages and disadvantages. Some of the drawbacks of these shape memory archwires may be prevented with the application of shape memory polymers [151]. Superior esthetics improved elastic modulus, enhanced mechanical and chemical stability and ability of the archwire to self-adjust during the treatment make the application of this material more suitable for the patient and the practitioner. Endodontics is another field that may benefit from 4D printing. One of the most common complications during root canal therapy is instrument separation. The ability of shape memory metals to adapt to root canal curvatures is an important asset that makes them competitive as a replacement of traditional nickel titanium rotary instruments in the future [12,232].

In prosthodontics, permanent seating of bridge pontics while the surrounding hard and soft tissues are undergoing changes may result in suboptimal fitting of the pontics. Application of shape memory polymers in these cases may be beneficial because of the self-adjustment capability of the material, which helps to minimize future complications. Full or partial removable dentures are often problematic for the patient because of bone resorption that accompanies tooth removal, as well as forces applied over the bony ridge by the denture. As a result, denture relining is a common procedure. The use of shape memory polymers for 4D printing of dentures may improve the ability of these dentures to adapt to the altering hard and soft tissue profiles. In implantology, titanium alloys are the most common materials used for fabricating oral implants. Although there are many benefits associated with titanium alloys, hypersensitivity and surface degradation have been reported. Replacement of titanium alloys with shape memory polymers may result in improved biocompatibility and better osseointegration [164]. Shape-memory polymers are advantageous by well filling the root canals, leading to prevent reinfections, biofilm formation, and periradicular diseases. They exhibit the natural tooth like expansion and contraction to fill the gap at the interface, decreasing the microleakage as well. Shape-memory polymers also produce biocompatible degradation products—they are biocompatible—as well as not produce metal ions which triggering chronic diseases, in comparison with metals. By the features far outweighing other types of materials, the shape-memory polymers show great stability and much more effectiveness, at least here in dentistry [233].

## 6. Conclusion and outlook

Three dimensional printing has been employed for rapid prototyping of different objects and equipments. Over the last decade, 3D printing technology has expanded rapidly across different medical sectors, including dentistry. New feed materials and printing methods are being developed to reduce design time, expedite fabrication and improve performance. Quality traits such as particle size, shape, distribution of the feed material powder as well as density, flowability and viscosity of the polymer directly affect the mechanical performance of the final product [234–237]. Viscosity-dependent flowability of polymers and metal allays is a challenge in 3D printing. Clinical dentistry requires fabrication of custom dental devices that differs from one patient to another. Dentistry may be the final industry for rapid prototyping technology to move into because it demands precise manufacturing and accuracy.

The advantages of 3D printing in dentistry have already been mentioned. It is important to point out that there are several aspects of 3D printing that require enhancement, such as improvement in surface quality, mechanical properties and dimensional accuracy. Printing products with better surface quality is directly related to the processing technique, the thickness of each layer and the depth of polymerization. The mechanical properties of the printed product are associated with porosity and surface quality. The porosity of a material depends on the particle size and processing temperature; material infiltration has been proposed to overcome this obstacle. Shrinkage of printing materials during manufacturing may result in significant reduction of mechanical properties. Dimensional accuracy of the products printed by 3D printers varies with the use of printing techniques and has a significant impact on the passive fit of dental prostheses.

**Acknowledgment**

P. M. and V. M. acknowledge funding from the European Horizon 2020 Research and Innovation Programme under Grant Agreement No 899349 (5D NanoPrinting).

**Conflict of Interest**

The authors declare no conflict of interest.